\documentclass[11pt,preprint]{aastex}
\usepackage{natbib}
\begin{document}

\shorttitle{Dependence of Disk and Host Star Masses}
\shortauthors{Andrews et al.}

\title{The Mass Dependence Between Protoplanetary Disks and their Stellar Hosts}

\author{Sean M. Andrews, 
        Katherine A. Rosenfeld, 
        Adam L. Kraus, and 
        David J. Wilner}

\affil{Harvard-Smithsonian Center for Astrophysics, 60 Garden Street, Cambridge, MA 02138, USA}

\email{sandrews@cfa.harvard.edu}

\begin{abstract}
We present a substantial extension of the millimeter-wave continuum photometry 
catalog for circumstellar dust disks in the Taurus star-forming region, 
based on a new ``snapshot" $\lambda = 1.3$\,mm survey with the Submillimeter 
Array.  Combining these new data with measurements in the literature, we 
construct a mm-wave luminosity distribution, $f(L_{\rm mm})$, for Class II 
disks that is statistically complete for stellar hosts with spectral types 
earlier than M8.5 and has a 3-$\sigma$ depth of roughly 3\,mJy.  The resulting 
census eliminates a longstanding selection bias against disks with late-type 
hosts, and thereby demonstrates that there is a strong correlation between 
$L_{\rm mm}$ and the host spectral type.  By translating the locations of 
individual stars in the Hertzsprung-Russell diagram into masses and ages, and 
adopting a simple conversion between $L_{\rm mm}$ and the disk mass, $M_d$, we 
confirm that this correlation corresponds to a statistically robust 
relationship between the masses of dust disks and the stars that host them.  A 
Bayesian regression technique is used to characterize these relationships in 
the presence of measurement errors, data censoring, and significant intrinsic 
scatter: the best-fit results indicate a typical 1.3\,mm flux density of 
$\sim$25\,mJy for 1\,$M_{\odot}$ hosts and a power-law scaling $L_{\rm mm} 
\propto M_{\ast}^{1.5-2.0}$.  We suggest that a reasonable treatment of dust 
temperature in the conversion from $L_{\rm mm}$ to $M_d$ favors an inherently 
linear $M_d \propto M_{\ast}$ scaling, with a typical disk-to-star mass ratio 
of $\sim$0.2--0.6\%.  The measured RMS dispersion around this regression curve 
is $\pm$0.7\,dex, suggesting that the combined effects of diverse evolutionary 
states, dust opacities, and temperatures in these disks imprint a FWHM range of 
a factor of $\sim$40 on the inferred $M_d$ (or $L_{\rm mm}$) at any given host 
mass.  We argue that this relationship between $M_d$ and $M_{\ast}$ likely 
represents the origin of the inferred correlation between giant planet 
frequency and host star mass in the exoplanet population, and provides some 
basic support for the core accretion model for planet formation.  Moreover, we 
caution that the effects of incompleteness and selection bias must be 
considered in comparative studies of disk evolution, and illustrate that fact 
with statistical comparisons of $f(L_{\rm mm})$ between the Taurus catalog 
presented here and incomplete subsamples in the Ophiuchus, IC 348, and Upper 
Sco young clusters.  
\end{abstract}

\section{Introduction} 

Planetary systems are forged from the gas-rich, dusty disks that orbit young 
stars.  The physical mechanisms related to that formation process are 
extraordinarily complex, so there is considerable ambiguity (and lively 
theoretical debate) on the dominant pathways and pitfalls involved in producing
the planets in our solar system and the larger exoplanet population from their
ancestral disks.  Yet despite all that uncertainty, theoretical models agree 
that the overall efficiency of the process depends strongly on the amount of 
raw material available in the circumstellar disk.  Therefore, in the context of 
planet formation, {\it mass} is a fundamental disk property.  The distribution 
of disk masses, $f(M_d)$, has been invoked as a key factor in accounting for 
the diverse demographic properties of the exoplanets 
\citep[e.g.,][]{ida04,ida08,alibert05,mordasini09}, and scaling trends between 
$M_d$ and other star/disk properties are speculated to generate some of the 
basic relations that are now being identified from observations of the 
exoplanet population.  The most notable example of the latter is the 
correlation between the giant planet frequency and stellar host mass 
($M_{\ast}$) noted by \citet{johnson07}, which was predicted theoretically in 
planet formation models that implicitly assumed there is a fundamental scaling 
relationship between $M_d$ and $M_{\ast}$ 
\citep[e.g.,][]{laughlin04,ida05,kornet06,alibert11}.

The best available observational diagnostic of $M_d$ comes from the thermal 
continuum emission generated by cool dust grains at (sub)millimeter/radio 
wavelengths \citep{beckwith90}.  In this part of the spectrum, the dust 
emission is optically thin over most of the disk volume: therefore, the mm-wave 
luminosity is directly proportional to the product of the total mass and 
(average) temperature of the dust, weighted by the grain emissivity -- $L_{\nu} 
\propto \kappa_{\nu} B_{\nu}(T) M_d$.  With this simple relation (and some 
reasonable assumptions for $\kappa_{\nu}$ and the gas-to-dust ratio), mm-wave 
photometry surveys can be used to construct a disk mass distribution
\citep{aw05,aw07b}, and study how it varies as a function of time 
\citep{carpenter05,lee11,mathews12}, environment \citep{mann09,mann10}, or host 
mass \citep[e.g.,][]{scholz06,schaefer09}.  Previous work has suggested that 
disk masses are substantially diminished by advanced age (by $\sim$3-5\,Myr) or 
proximity to massive (OB-type) stellar neighbors, and are perhaps intrinsically 
lower for later type (M) stellar hosts.  In principle, such results hold clues 
to the nature of disk evolution and some of the key initial conditions relevant 
to the planet formation process.

However, the quantitative inferences of those studies should be regarded with 
caution.  In practice, this work relies on comparisons relative to a 
``reference" $f(M_d)$, constructed from a deep mm-wave photometry survey of 
disks in the Taurus region \citep{aw05}.  Because of its proximity ($d \approx 
140$\,pc) and youth ($\sim$2\,Myr), Taurus receives 
considerable (some might say disproportionate) attention from astronomers.  
However, it comprises the most complete and best-characterized young cluster 
available, and therefore is the most desirable stellar population to use in 
building a reference $f(M_d)$ for the comparative work noted above.  
Unfortunately, it is not generally appreciated that such comparisons are 
problematic because the reference census of \citet{aw05} is {\it incomplete} 
and {\it biased} against low-mass stellar hosts.  In fact, nearly half of the M 
stars with disks in Taurus had regrettably never been observed at millimeter 
wavelengths.  Since the M stars represent the peak of the stellar mass function 
and a wide range in $M_{\ast}$ (a factor of $\sim$10), a complete census of 
their disk masses is vital in efforts to develop a more appropriate reference 
$f(M_d)$ and to facilitate a robust search for any relationship between disk 
and star masses.

In this article, we present an extension of the Taurus disk mass survey based 
on ``snapshot" observations with the Submillimeter Array that is complete for 
known Class II sources with spectral types earlier than M8.5, and has a mm-wave 
luminosity sensitivity comparable to the \citet{aw05} study.  In Section 2, we 
describe the extension of the sample as well as the new millimeter continuum 
observations and their calibration.  In Section 3, we construct an updated 
distribution of mm-wave flux densities ($\propto M_d$) for Taurus disks, 
consider how it depends on various environmental and evolutionary factors, and 
directly explore the basis for a relationship between $M_d$ and $M_{\ast}$.  
The results are discussed in Section 4 in the contexts of practical 
consequences for future comparative observational work, implications for 
theoretical models of planet formation, and potential demographic connections 
in the populations of circumstellar disks and exoplanets.  A summary of the key 
conclusions of this work is provided in Section 5, and an extensive data 
catalog is made available in an Appendix (and through a web-accessible 
electronic database).

\section{Sample Selection and New Observations}

The Taurus-Auriga region includes a few hundred young, low-mass stars 
interspersed among small groups of dark clouds, spanning roughly 100 square 
degrees on the sky \citep[e.g.,][]{kenyon08}.  Historical spectroscopic 
parallax estimates have favored a mean distance of $d \approx 140\pm20$\,pc to 
the association, in good agreement with a few precise, individual trigonometric 
parallaxes from VLBI measurements \citep{loinard07,torres09,torres12}.  
Although a distributed population of Class III (young, but presumably 
disk-less) sources may still remain elusive 
\citep[e.g.,][]{herbig78,neuhauser95,slesnick06}, there is a robust, vetted, 
and effectively complete list of Class II (those with infrared excesses 
indicative of disks) members available thanks to the deep, wide coverage 
afforded by various {\it Spitzer} photometry surveys 
\citep{luhman10,rebull10}.  The infrared-selected \citeauthor{luhman10}~catalog 
of Tau-Aur Class II sources is used as the basis of our sample: it includes 179 
young stellar ``systems" composed of 227 (known) individual stars with spectral 
types from B8 to M9 \citep[e.g.,][]{luhman00,luhman04,kraus11}.  

\begin{figure}[t!]
\epsscale{0.5}
\plotone{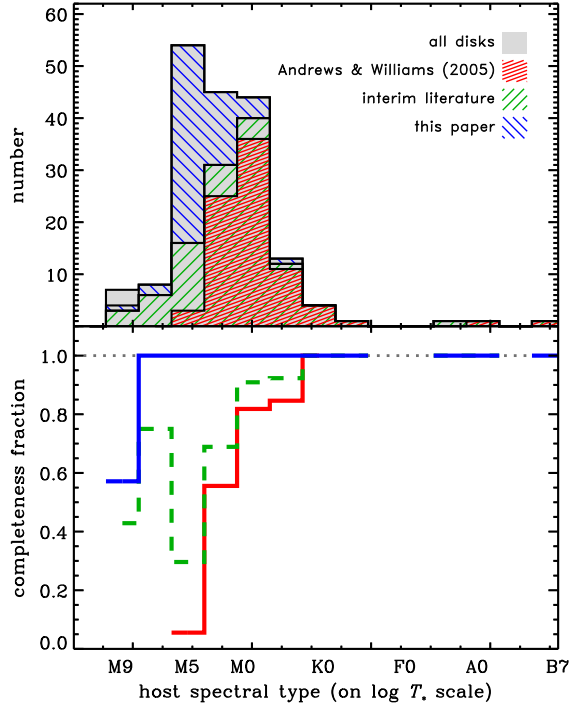}
\figcaption{({\it top}) The distribution of stellar host spectral types (on a 
$\log{\,T_{\ast}}$ scale; see Appendix A) for all Taurus Class II members ({\it 
gray} background), those in the \citet{aw05} mm-wave photometry catalog ({\it 
red}), plus new measurements since that survey ({\it green hatched}), and 
finally including those that have been observed here ({\it blue hatched}).  The 
new data cover most of the M-type hosts.  ({\it bottom}) The fractional 
completeness of the mm-wave photometry in the same spectral type bins.  The 
extended sample presented here is complete for types earlier than M8.5.  
\label{fig:completeness}}
\end{figure}

Of those 179 systems in the full sample, only 82 ($\sim$46\%) were included in 
the \citet{aw05} (sub)mm continuum photometry catalog (or its precursors).  The 
incompleteness of that catalog is due to a strong bias with respect to the 
spectral type of the host star, such that most of the M type stars were not 
observed: measurements were available for 31/50 (62\%) of the M0-M2 hosts, 
16/45 (36\%) for M2-M5, and {\it none} of those later than M5.\footnote{For 
historical purposes, it is worth pointing out some of the underlying 
circumstances that resulted in this selection bias.  First, the \citet{aw05} 
survey was largely conducted in 2003-2004, before the full population of late 
type Class II sources had been identified or confirmed through {\it Spitzer} 
infrared photometry surveys.  And second, even had a full sample been vetted, 
the long-term survey goals of that project would still have been cut short by 
the (premature) decomissioning of the primary instrument used for the 
photometry (SCUBA on the JCMT).}  A clear demonstration of this selection bias 
is shown in Figure \ref{fig:completeness}.  Since the \citet{aw05} catalog was 
published, an additional 34 sources have been observed at (sub)mm wavelengths, 
including a number of M-type hosts \citep[e.g.,][]{scholz06,schaefer09}.  
However, Figure \ref{fig:completeness} demonstrates that even this updated 
catalog remains highly incomplete for late-type hosts.  To remedy that issue, 
we identified 60 of those remaining sources and designed a survey to measure 
their mm-wave continuum emission.  Combined with the previous measurements, 
this new sample of 176 sources is {\it complete} for Class II members with host 
spectral types earlier than M8.5 (and $\sim$50\%\ complete for later types).  

We observed 52 new targets with the Submillimeter Array \citep[SMA;][]{ho04} in 
its compact configuration (8-70\,m baselines) over six tracks from 2012 
November to 2013 February.  The data were collected in a ``snapshot" observing 
mode, where short integrations on $\sim$5-11 individual targets were 
interspersed throughout each track, along with brief visits to the nearby 
quasar 3C 111.  Additional observations of Uranus, Callisto, 3C 279, and 3C 84 
were conducted for calibration purposes when the science targets were at low 
elevations.  The total integration time per target was $\sim$45\,minutes.  
Although this snapshot approach degrades the image fidelity, it facilitates 
robust flux density measurements for a large number of targets in a short 
period of time.  The SMA double-sideband receivers were tuned to a local 
oscillator (LO) frequency of 225.5\,GHz (1.33\,mm), and the correlator was 
configured to process two IF bands that spanned $\pm$4--8\,GHz from the LO with 
$48\times108$\,MHz spectral chunks (each with 32 individual channels) per 
sideband.  The observing was performed in good weather conditions, with 
precipitable water vapor levels in the 1-2\,mm range.  An additional 8 targets 
were found to have useful data in the SMA archive, and are included here.

The data were calibrated with the {\tt MIR} software package.  The spectral 
response of the system was corrected using observations of 3C 279 and 3C 84, 
and the visibility amplitude scale was bootstrapped from the observations of 
Uranus or Callisto, depending on their availability.  The absolute amplitude 
scale has a systematic uncertainty of $\sim$10\%.  Antenna-based complex gain 
variations from instrumental and atmospheric effects were removed by 
referencing to regular observations of 3C 111.  The visibility spectra for each 
science target were averaged into a composite wideband continuum dataset for 
each IF band/sideband pair and then concatenated.  The visibilities were 
Fourier inverted, deconvolved with the {\tt CLEAN} algorithm, and then restored 
with the synthesized beam using the {\tt MIRIAD} package.  The resulting 
continuum emission maps have RMS noise levels of 0.5--2.1\,mJy beam$^{-1}$.  
Flux densities were measured both from these maps and directly from the 
visibilities (both methods are in agreement in all cases): when no emission is 
detected, we report upper limits that correspond to 3$\times$ the RMS noise 
level in the corresponding map.  Table \ref{tab:obs} lists the flux densities 
or 3-$\sigma$ upper limits for the 60 new targets provided by this survey.

\section{Analysis and Results} 

Combined with previous measurements in the literature, the new observations 
described above allow us to construct a deep mm-wave photometry catalog of 
Class II disks that is complete for stellar host spectral types earlier than 
M8.5.  In this section, we generate and analyze an updated set of mm-wave 
continuum luminosity distributions (\S 3.1), and investigate the potential 
dependence of the emission from these dust disks (i.e., their masses) on the 
masses of their stellar hosts (\S 3.2).

\subsection{Millimeter-wave Luminosity Distributions}

Before starting any analysis of the full sample, the diverse set of literature 
photometry measurements needs to be homogenized.  The extension catalog 
described in \S 2 was observed at $\lambda = 1.33$\,mm, but the literature 
catalogs include photometry (or upper limits) for each source at wavelengths of 
0.86--0.88 and/or 1.2--1.4\,mm.  Often, the sources with measurements in both 
wavelength ranges have one that is significantly more reliable or useful due to 
its higher signal-to-noise ratio or intrinsically deeper limit.  These issues 
can be mitigated by using all of the available data to generate representative 
flux densities (or upper limits) from an extrapolation to a ``reference" 
frequency.  To facilitate comparisons with future work, two reference 
frequencies are adopted here based on our expectations of common observing 
setups for surveys with the Atacama Large Millimeter/Submillimeter Array 
(ALMA): $\nu_{\rm ref} = 225$\,GHz (1.3\,mm) and 338\,GHz (0.89\,mm).  The 
former corresponds to an ALMA configuration that simultaneously covers the 
$^{12}$CO/$^{13}$CO/C$^{18}$O $J$=2$-$1 lines, while the latter includes the 
$J$=3$-$2 transitions of $^{12}$CO/$^{13}$CO and avoids regions of poorer 
atmospheric transmission.  For the 60 sources in the full catalog with $\ge$2 
detections in the 0.7--3\,mm wavelength range, we determined flux densities at 
the reference frequencies from power-law fits, where $F_{\nu} \propto 
\nu^{\alpha}$.  In all other cases, we applied an extrapolation based on a 
weighted average of those fit results, with an effective index $\langle \alpha 
\rangle = 2.4\pm0.5$.  The uncertainties on the homogenized flux densities 
include the uncertainty on the index $\alpha$.  For sources with no available 
detections, the intrinsically deeper observation is used to extrapolate upper 
limits at the reference frequencies.  Since many more sources in the full 
sample have measurements at (or near) 1.3\,mm, our analysis and results are 
discussed in the context of this wavelength.  A comment on notation: in this 
article, we use the terms mm-wave ``flux density" ($F_{\rm mm}$) and 
``luminosity" ($L_{\rm mm}$) interchangably, but prefer to employ the 
standard units (Jy) of the former for ease of use in literature 
comparisons.\footnote{Note that $L_{\rm mm}/L_{\odot} \approx 0.0014 (F_{\rm 
mm}/{\rm Jy})$ at 1.3\,mm, or $\approx 0.0021 (F_{\rm mm}/{\rm Jy})$ at 
0.89\,mm, for a distance $d = 140$\,pc.}

Next, we adopt an approach for the assignment of emission to individual 
components of multiple star systems.  The recent component-resolved imaging 
survey by \citet{harris12} is used for brighter systems with separations 
$>$0\farcs3.  When no component-resolved measurements are available, $F_{\rm 
mm}$ values are assigned based on the projected separation between a pair of 
stars.  For close pairs ($\le$0\farcs1, roughly 15\,AU), dynamical simulations 
of star-disk interactions suggest that individual disks are unlikely to survive 
\citep[e.g.,][]{artymowicz94}.  In those cases, we associate the emission (or 
lack thereof) with a circumbinary disk around both components.  For undetected 
systems with wider separations, the measured upper limit is assigned to each 
individual component.  Finally, there are 8 multiple systems\footnote{FO Tau, 
FS Tau, XZ Tau, GH Tau, IS Tau, Haro 6-28, GN Tau, and V955 Tau.} (containing 
16 stars) that are detected but not yet resolved at mm wavelengths.  In those 
cases, we assume that all of the emission is associated with the primary and 
assign the secondaries an upper limit corresponding to 3$\times$ the quoted RMS 
noise level for the system.  For this last scenario, our motivation comes from 
two considerations.  First, \citet{harris12} found that the primary component 
always dominates the $F_{\rm mm}$ budget for binary pairs, and is often the 
{\it only} component of a multiple system with any mm-wave emission.  And 
second, the multiplicity census for late-type stellar hosts is likely 
incomplete, meaning that we may already be de facto assigning $F_{\rm mm}$ in 
this way for some sources that we assume are single, but may actually have 
faint companions.  Overall, this assumption applies to few enough sources that 
it is effectively inconsequential in the analysis that follows: alternative 
assumptions -- e.g., that the emission is distributed equally to each component 
-- do not significantly impact any of the results.  

Based on the homogenization and emission assignments outlined above, a catalog 
of $F_{\rm mm}$ values at both reference frequencies for the full sample of 
Taurus Class II members is provided in Table \ref{tab:Fmm}.  For reference 
purposes, the original photometric measurements used to determine these 
representative $F_{\rm mm}$ are available for each individual source in 
electronic format in Appendix A.

\begin{figure}[t!]
\epsscale{1.1}
\plottwo{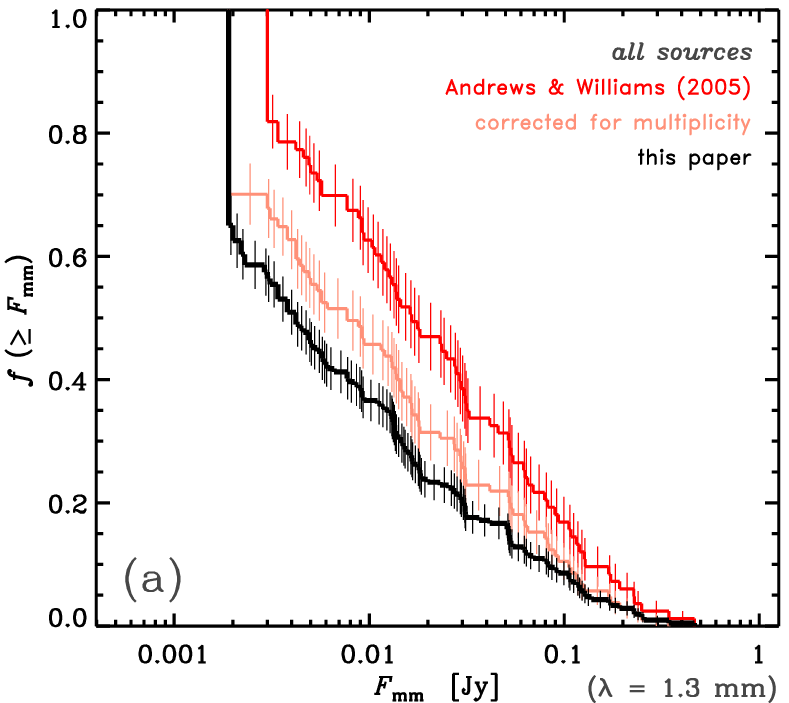}{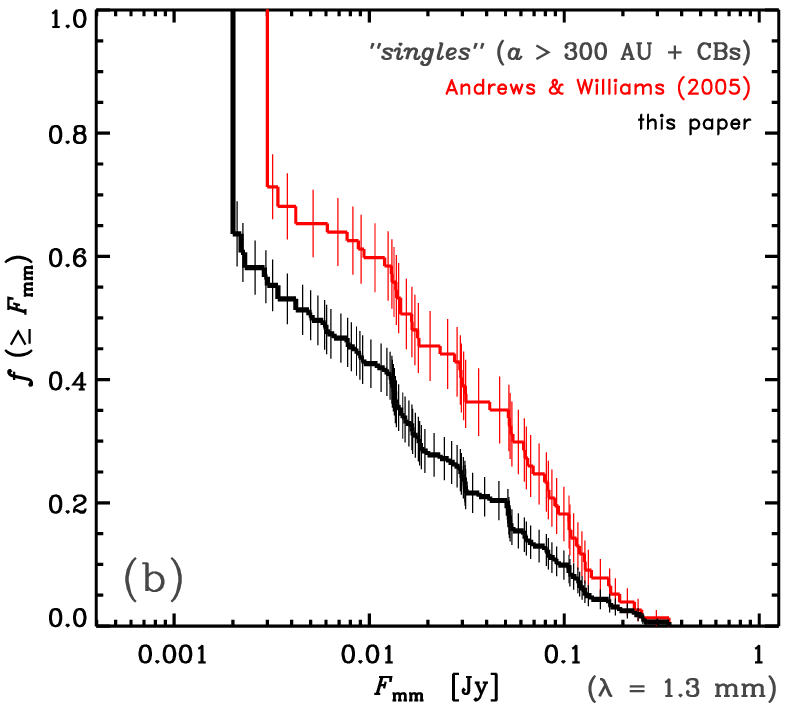}
\figcaption{(a) The cumulative mm-wave luminosity distributions for the 
complete sample presented here ({\it black}) and the original \citet{aw05} 
catalog as published ({\it red}) and corrected for multiplicity ({\it pink}).  
The complete sample has systematically lower luminosities, due to selection 
biases in previous catalogs.  (b) The same cumulative distributions, but only 
for sources where the influence of a companion on the disk properties is 
minimal: single stars, close pairs with circumbinary disks, and wide binaries.  
The difference between the two curves is entirely due to selection effects 
related to the distribution of host stars in the two samples. \label{fig:cdfs}}
\end{figure}

The catalog in Table \ref{tab:Fmm} is used to generate the cumulative 
distribution of mm-wave luminosities, $f(\ge$$F_{\rm mm}$), shown in Figure 
\ref{fig:cdfs}(a) ({\it black curve}).  This distribution corresponds to a 
statistically complete sample of Class II disks in Taurus with host spectral 
types earlier than M8.5, and was constructed using the Kaplan-Meier 
product-limit estimator to properly account for censored measurements (upper 
limits on $F_{\rm mm}$) using the formalism described by \citet{feigelson85}.  
Its shape can be described reasonably well using a log-normal function with a 
mean of 4\,mJy (the median $F_{\rm mm}$ is the same) and a large standard 
deviation, 0.9\,dex (at the alternative reference wavelength of 0.89\,mm, the 
mean is 11\,mJy and the standard deviation is also $\sim$0.9\,dex).  

This cumulative luminosity distribution is dramatically different than the one 
derived from the \citet{aw05} catalog; the latter is also shown in Figure 
\ref{fig:cdfs}(a) ({\it red curve}), updated to use the homogenized $F_{\rm 
mm}$ values.  Some of that discrepancy is related to the proper assignment of 
emission in multiple systems: \citeauthor{aw05} (necessarily) treated these as 
composite systems.  After applying a correction based on the accounting system 
described above ({\it pink curve}), we find that the overall shapes of the 
$f$($\ge$$F_{\rm mm})$ distributions are similar, but the complete sample is 
preferentially shifted to $\sim$50\%\ lower luminosities (for $F_{\rm mm} < 
0.1$\,Jy).  The blanket inclusivity of Figure \ref{fig:cdfs}(a) obscures how 
notable the differences are between the old and new catalogs.  Figure 
\ref{fig:cdfs}(b) compares the $f(\ge$$F_{\rm mm})$ for sources where 
multiplicity is not expected to substantially diminish the mm-wave emission -- 
for single stars, close pairs with circumbinary disks, and wide binaries with 
projected separations $a > 300$\,AU \citep{harris12} -- and highlights a large 
discrepancy that more faithfully reflects the impact of incompleteness in the 
\citet{aw05} sample.  In short, a comparison between the mm-wave luminosity 
distributions for the complete sample and the standard \citet{aw05} reference 
catalog confirms that the latter is demonstrably biased toward brighter 
sources.  Since the primary difference between the two samples was the addition 
here of many new targets with late-type stellar hosts (see 
Fig.~\ref{fig:completeness}), it makes sense to suspect that such sources 
exhibit preferentially fainter mm-wave emission.

\begin{figure}[t!]
\epsscale{1.1}
\plottwo{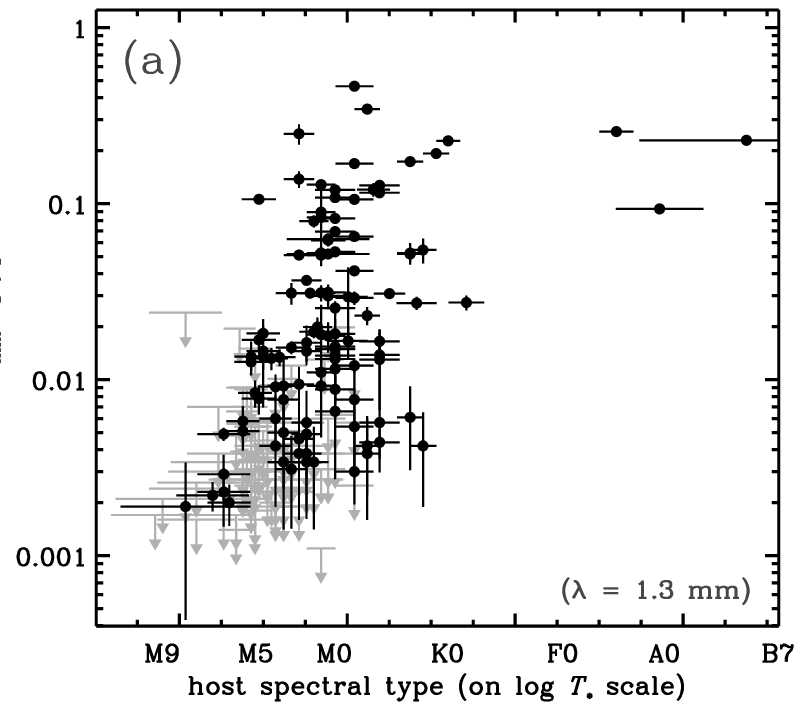}{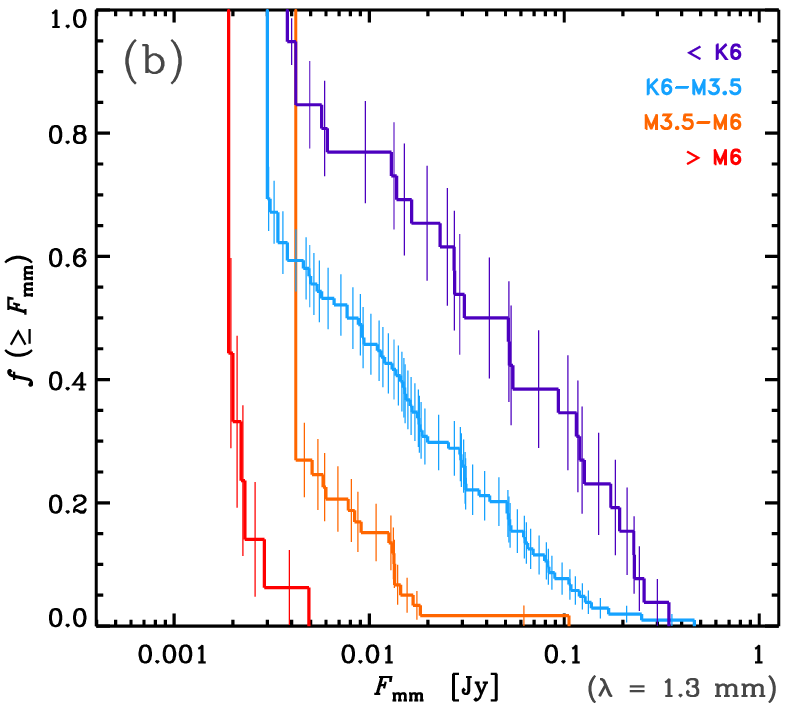}
\figcaption{(a) The dependence of the mm-wave luminosities from dust disks 
(3\,$\sigma$ upper limits are marked as {\it grey} arrows) on the spectral 
types of their stellar hosts: the abscissa axis represents a logarithmic 
effective temperature scale (see text for details).  Statistical tests confirm 
that there is a clear correlation between these two variables.  (b) An 
illustration of the same relationship, through a comparison of cumulative 
luminosity distributions for different spectral type bins.  The mm-wave 
luminosities are systematically larger for disks around stellar hosts with 
earlier spectral types. \label{fig:Fmm_spt}}
\end{figure}

That simple inference is directly examined for the full sample catalog in 
Figure \ref{fig:Fmm_spt}(a), which plots $F_{\rm mm}$ as a function of the host 
spectral type.  The abscissa values are marked on a logarithmic effective 
temperature ($T_{\ast}$) scale that assumes the standard correspondence with 
spectral types earlier than M1 \citep{schmidt-kaler82,straizys92} and the 
updated $T_{\ast}$ scale advocated by \citet{luhman99} for later types (as in 
Fig.~\ref{fig:completeness}; see Appendix B for more details).  The behavior in 
this plot clearly demonstrates that later type stellar hosts harbor disks with 
systematically lower emission levels, and have an overall lower detection rate 
compared to their counterparts with earlier types.  The standard correlation 
tests for censored datasets described by \citet{isobe86} -- the Cox 
proportional hazard, generalized Kendall rank, and generalized Spearman rank 
tests -- all verify that the null hypothesis is ruled out with very high 
confidence (i.e., the probability that there is no correlation between $F_{\rm 
mm}$ and $T_{\ast}$ is very low, $p_{\emptyset} < 10^{-8}$).  Figure 
\ref{fig:Fmm_spt}(b) renders the correlation more visually recognizable by 
splitting the cumulative luminosity distribution into spectral type bins, and 
thereby showing a clear increasing progression of $F_{\rm mm}$ for earlier 
spectral types.  

The identification of this correlation between $L_{\rm mm}$ and $T_{\ast}$ is 
possible only because the sample presented here is complete over a large range 
of host spectral types.  That new capability illustrates clearly why it is so 
important to obtain complete samples without selection biases (or to properly 
account for them; see \S 4) in demographic studies of disk populations.  But 
perhaps more important, it provides some tantalizing evidence for an intrinsic 
relationship between the masses of disks and their stellar hosts: $L_{\rm mm}$ 
is roughly proportional to $M_d$, and $T_{\ast}$ (i.e., the spectral type) 
scales with $M_{\ast}$ over a wide range of young star masses.  Nonetheless, 
there is a large amount of scatter both in the data and in the intrinsic 
relationships between these {\it observed} properties, \{$L_{\rm mm}$, spectral 
type\}, and their {\it derived} analogs, \{$M_d$, $M_{\ast}$\}.  The following 
section aims to map the conversion between the two and explore the basis for a 
relationship between $M_d$ and $M_{\ast}$ more explicitly.

\subsection{Dependence on Stellar Host Masses}

\subsubsection{Derivation of $M_{\ast}$ Estimates}

In our effort to derive these fundamental stellar properties, we use the 
Bayesian inference approach first derived by \citet{jorgensen05}, and later 
developed by \citet{gennaro12}.   For each individual star, the goal is to 
determine the joint likelihood function $\mathcal{L}(M_{\ast}, t_{\ast} | 
T_{\ast}, L_{\ast})$, which characterizes the desired model properties (mass 
and age) conditioned on a measurement of its temperature and luminosity.  
However, a determination of \{$T_{\ast}$, $L_{\ast}$\} and their associated 
uncertainties from directly observable quantities is not trivial.  Here, we 
adopt a simple scaling relation with the spectral classification to assign 
$T_{\ast}$ (see \S 3.1) and then use a stochastic optimization fitting method 
to determine $L_{\ast}$ (and an extinction, $A_V$) by matching scaled and 
reddened spectral templates of stellar photosphere models to the broadband 
optical/near-infrared spectral energy distribution (SED).  Since this effort is 
not intended to be the focus here, we relegate a detailed explanation of this 
process and a compilation of the results to Appendix B.  

\begin{figure}[t!]
\epsscale{1.0}
\plotone{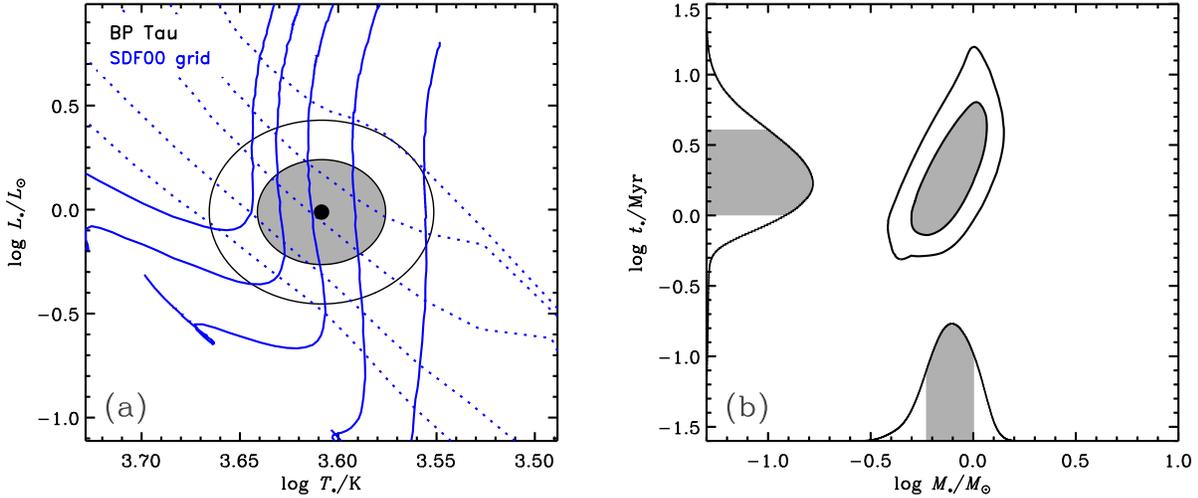}
\figcaption{An illustration of the method used to estimate stellar masses and 
ages, for the case of BP Tau and the SDF00 models.  (a) The local H-R diagram 
with the SDF00 tracks for $M_{\ast} = 0.4$, 0.6, 0.8, 1.0, and 1.2\,$M_{\odot}$ 
(solid curves, from right to left) and isochrones for $t_{\ast} = 0.5$, 1, 2, 
5, and 10\,Myr (dotted curves, from top to bottom) overlaid.  The black dot 
represents the measured \{$\log{\,T_{\ast}}$, $\log{\,L_{\ast}}$\} values, and 
the contours around it mark the 68 and 95\%\ confidence intervals of the 
likelihood function in Eq.~(1).  (b) The likelihood function mapped onto 
\{$\log{\,M_{\ast}}$, $\log{\,t_{\ast}}$\}-space.  The marginal probability 
density functions for each parameter are shown along their respective axes.  
The best-fit parameters and their uncertainties listed in Table \ref{tab:stars} 
are determined from the peaks of those distributions and the 68\%\ confidence 
intervals (in shaded gray), respectively.  \label{fig:HRhowto}}
\end{figure}

The underlying ``model" involved in this technique is a grid of pre-main 
sequence (pre-MS) stellar evolution calculations.  To capture the intrinsic 
uncertainties in (or disagreements between) these grids themselves 
\citep[e.g., see][]{hillenbrand04}, we consider three different sets of model 
calculations: \citet[][hereafter DM97]{dantona97} with the updated prescription 
for deuterium burning \citep{dantona98}, \citet[][BCAH98]{baraffe98}, and 
\citet[][SDF00]{siess00}.  Each grid is used to construct a finely interpolated 
(discretized) model H-R diagram, with coordinates \{$\hat{T}$, $\hat{L}$\}.  
For each individual star and pre-MS model grid, a conditional likelihood 
function is then defined and evaluated for each location in the model H-R 
diagram: 
\begin{equation}
\mathcal{L}(\hat{T}, \hat{L} | T_{\ast}, L_{\ast}) = \frac{1}{2\pi\sigma_T\sigma_L} \exp  \left( -\frac{1}{2} \left[\frac{(T_{\ast}-\hat{T})^2}{\sigma_T^2} + \frac{(L_{\ast}-\hat{L})^2}{\sigma_L^2} \right] \right),
\end{equation}
where $\sigma_T$ and $\sigma_L$ are the measured uncertainties (assumed to be 
normally distributed) associated with $T_{\ast}$ and $L_{\ast}$, respectively 
(we implicitly assume uniform priors on the model parameters, \{$\hat{T}$, 
$\hat{L}$\}, in this calculation).  
A distance uncertainty of $\pm$20\,pc is included in $\sigma_L$, added in 
quadrature to the nominal uncertainty (see Appendix B).  The pre-MS model grids 
provide a direct, one-to-one mapping \{$\hat{T}$, $\hat{L}\} \mapsto 
\{M_{\ast}, t_{\ast}$\}, and therefore a corresponding mapping 
$\mathcal{L}(\hat{T}, \hat{L} | T_{\ast}, L_{\ast}) \mapsto 
\mathcal{L}(M_{\ast}, t_{\ast} | T_{\ast}, L_{\ast})$.\footnote{In practice, 
all of the analysis described here is performed on logarithmic variables to 
facilitate a more straightforward numerical evaluation of the likelihood grid 
over the wide ranges spanned by the parameters of interest.  For the sake of 
clarity in notation, we have omitted the $\log{X}$ in favor of the simpler $X$ 
for each variable.}  The marginal probability density functions $p(M_{\ast} | 
T_{\ast}, L_{\ast})$ and $p(t_{\ast} | T_{\ast}, L_{\ast})$ can then be 
determined through an appropriate numerical integration of 
$\mathcal{L}(M_{\ast}, t_{\ast} | T_{\ast}, L_{\ast})$ over $t_{\ast}$ and 
$M_{\ast}$, respectively.  For reference, a graphical illustration of the 
approach is shown in Figure \ref{fig:HRhowto}.  

The method outlined above was used to determine the best-fit 
\{$\log{\,M_{\ast}}$, $\log{\,t_{\ast}}$\} (corresponding to the peaks of the 
marginal distributions for each parameter) and their associated uncertainties 
(the marginalized 68\%\ confidence intervals) for each individual star and each 
of the three pre-MS model grids.  The results are compiled in Table 
\ref{tab:stars}.  It is worth explicitly pointing out that each pre-MS model 
grid has boundaries, and the inferences of \{$M_{\ast}, t_{\ast}$\} outside 
those boundaries should be considered highly uncertain.  This is a relatively 
minor concern for $M_{\ast} < 0.1$\,$M_{\odot}$ in the SDF00 models, but is a 
serious issue when $M_{\ast} > 1.4$\,$M_{\odot}$ or $t_{\ast} < 1$\,Myr in the 
BCAH98 models.  Whenever relevant in the following, we are sure to highlight 
these extrapolated quantities.  

\begin{figure}[t!]
\epsscale{1.1}
\plottwo{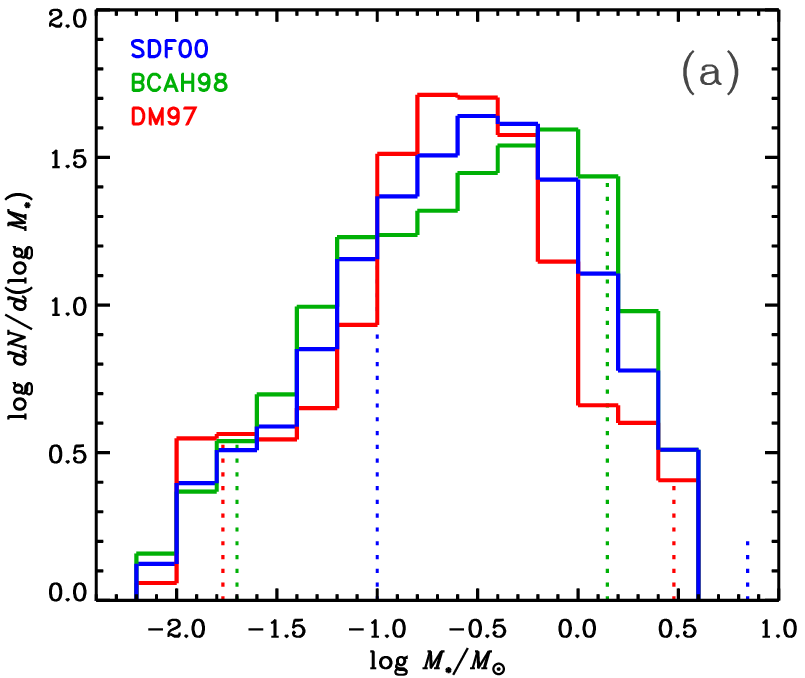}{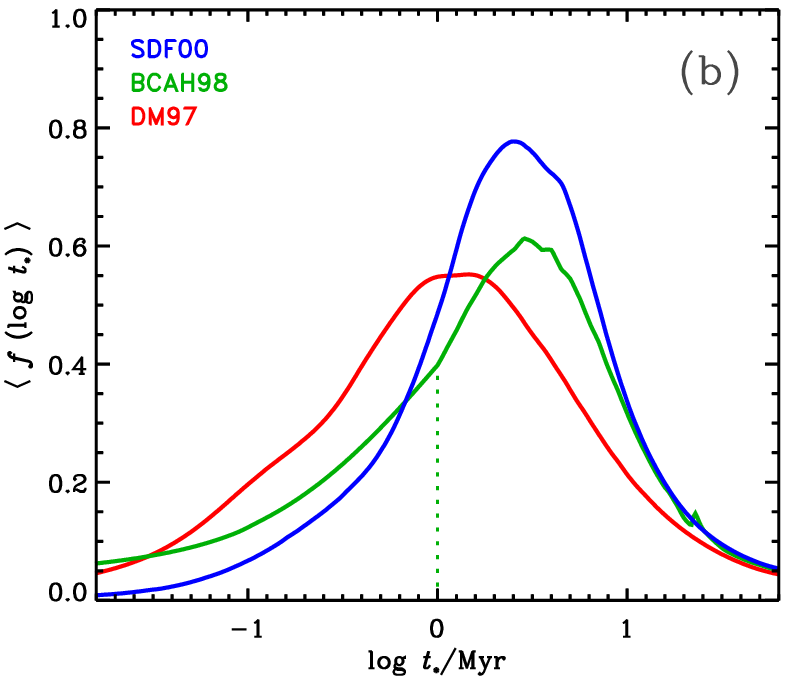}
\figcaption{(a) The ensemble stellar mass functions for the sample, constructed 
by summing the marginal $M_{\ast}$ probability density functions for individual 
sources and then integrating into discrete bins (thereby properly incorporating 
the uncertainty on each individual measurement).  Dotted vertical lines mark 
the boundaries of each pre-MS model grid: outside those lines, the $M_{\ast}$ 
values were determined from extrapolations.  (b) The ensemble mean age 
distributions for the sample, determined by averaging the marginal $t_{\ast}$ 
probability density functions for individual sources.  Ages less than the green 
dotted line for the BCAH98 models are extrapolations.  \label{fig:MTensemble}}
\end{figure}

Before moving on, it is worthwhile to comment on the stellar properties 
inferred for this catalog.  Figure \ref{fig:MTensemble}(a) displays the 
ensemble stellar mass functions, constructed by summing (and then binning) the 
marginal distributions, $p(M_{\ast} | T_{\ast}, L_{\ast})$, for each individual 
star in the sample.  The morphology is similar to previous measurements of the 
overall Taurus mass function \citep[e.g.,][]{luhman00,luhman04,briceno02}, 
although this version has the added benefit of explicitly incorporating the 
uncertainties for each contributing $M_{\ast}$ estimate.  Taken together, the 
three different pre-MS model grids predict mass functions for this sample that 
are generally consistent with one another.  A similar approach was employed in 
Figure \ref{fig:MTensemble}(b) to determine a mean age distribution for the 
sample, in the view of each pre-MS model grid.  The BCAH98 and SDF00 models 
both suggest an ensemble mean age of $\sim$2.5--3\,Myr for Taurus Class II 
sources; the DM97 models argue for a younger age, $\sim$1\,Myr.  This 
systematic offset between the isochronal ages is primarily related to the 
assumed initial conditions in the evolutionary model calculations 
\citep[e.g.,][]{stahler83,palla93}, as was noted in a similar comparison of 
ages inferred from the DM97 and other model grids for young stars in Orion 
\citep{hillenbrand97}.  However, despite this apparent discrepancy, the widths 
of these mean age distributions are very large -- 68\%\ of their areas are 
enclosed in a range of $\sim$0.2--8\,Myr -- suggesting reasonable agreement 
overall.

Although the three different pre-MS model grids do make different predictions 
for the stellar properties that correspond to each location in the H-R diagram, 
there is generally good agreement between them within the (substantial) 
uncertainties inferred with this technique.  We find a modest tendency for the 
BCAH98 models to predict higher masses: on average $\sim$0.2\,dex above the 
DM97 values and $\sim$0.1\,dex higher than SDF00, although the RMS dispersions 
in the residuals are of the same order and that progression is reversed over a 
narrow mass range ($M_{\ast} \approx 0.1$-0.2\,$M_{\odot}$).  These features 
are manifestations of structural differences between the models, tied to the 
resulting $\hat{T}$ values.  Dynamical mass estimates ($M_{\rm dyn}$) are 
available for $\sim$20 sources in this sample, derived either from gas disk 
rotation curves \citep{simon00,dutrey03,dutrey08,pietu07,schaefer09,tang12}, 
radial velocity monitoring of spectroscopic binaries \citep[plus some external 
inclination constraint; e.g.,][]{mathieu97,prato02}, or the reconstruction of 
astrometric orbits for close binaries \citep{tamazian02,schaefer06}.  A 
comparison of $M_{\rm dyn}$ and our estimates of $M_{\ast}$ shows a pattern 
similar to the inter-model comparisons: the BCAH98 models slightly over-predict 
$M_{\rm dyn}$ by $\sim$0.1\,dex, the SDF00 models marginally under-predict by 
$\sim$0.05\,dex, and the DM97 models under-predict by $\sim$0.2\,dex (the RMS 
dispersions of the residuals are $\sim$0.1\,dex).  However, it is worth noting 
that the range of $M_{\rm dyn}$ measurements available for Class II Taurus 
sources is still biased to solar-mass (or greater) systems.  Quantitatively, 
this level of mass disagreement is similar to what was found by 
\citet{hillenbrand04} for a larger, and primarily older, sample.  The stellar 
ages inferred here have considerably more (formal) uncertainty associated with 
them.  Again, on an individual basis the $t_{\ast}$ estimates from different 
models are consistent within the uncertainties, although the best-fit DM97 ages 
are systematically younger (a median shift of $\sim$0.3\,dex) compared to both 
the BCAH98 and SDF00 predictions.

\subsubsection{A Statistical Comparison of $M_d$ and $M_{\ast}$}

\begin{figure}[t!]
\epsscale{1.0}
\plotone{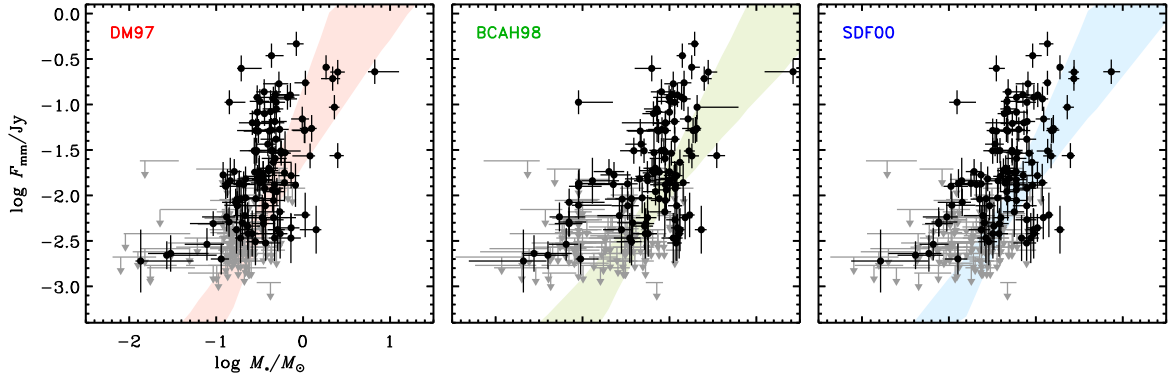}
\figcaption{A comparison of $L_{\rm mm}$ and $M_{\ast}$ for each set of pre-MS 
models (upper limits are shown as gray arrows).  The colored swaths mark 95\%\
confidence intervals on the $\log{L_{\rm mm}} \sim \log{M_{\ast}}$ relationship,
derived from a Bayesian linear regression analysis that accounts for censored
data.  \label{fig:mdvms}}
\end{figure}

\begin{figure}[t!]
\epsscale{1.0}
\plotone{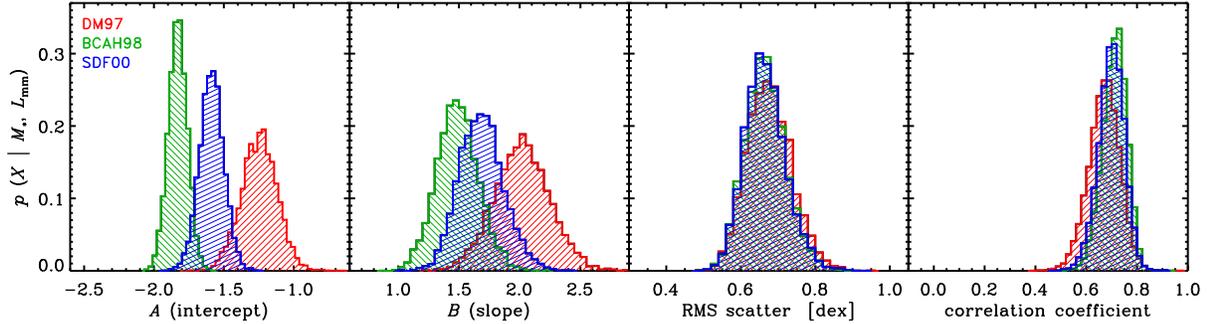}
\figcaption{Results from a Bayesian linear regression analysis of the
relationship between $\log{L_{\rm mm}}$ and $\log{M_{\ast}}$.  From left to 
right, the marginal posterior probability density functions for the intercept 
($A$), slope ($B$), dispersion around the regression, and corresponding 
correlation coefficient.  \label{fig:regress}}
\end{figure}

Now, using the $M_{\ast}$ values derived above (Table \ref{tab:stars}), we can 
make a direct mass comparison between disks and their stellar hosts.  First, we 
consider the relationship between the observational proxy for disk mass, 
$L_{\rm mm}$, and $M_{\ast}$.  The plots in Figure \ref{fig:mdvms} clearly 
indicate that the mm-wave luminosities from dust disks are highly correlated 
with their host star masses in the full Taurus Class II sample (210 
datapoints).  That visual exhibition of a strong relationship between $L_{\rm 
mm}$ and $M_{\ast}$ is quantitatively confirmed by correlation tests adapted 
for use on censored datasets \citep{isobe86}.  The Cox proportional hazard test 
suggests that the null hypothesis (no correlation) has a probability 
$p_{\emptyset} < 10^{-8}$ for each of the pre-MS models (with global $\chi_r^2$
values of 56, 73, and 72 for DM97, BCAH98, and SDF00, respectively).  Similarly
stringent limits on $p_{\emptyset}$ are derived from the generalized Kendall or
Spearman rank tests (with $z$-scores of 8.3, 9.1, and 9.1 and $\rho = 0.55$,
0.61, and 0.60 for each model grid and correlation test, respectively).  
Assuming an intrinsic power-law scaling between $L_{\rm mm}$ and $M_{\ast}$, 
this correlation is quantified with a linear regression analysis in the 
log--log plane, where $\log{(L_{\rm mm}/L_{\odot})} = A + 
B\log{(M_{\ast}/M_{\odot})}$, using the Bayesian methodology developed by 
\citet{kelly07} to properly account for the measurement uncertainties, data 
censoring, and substantial scatter along the correlation.  The results of that 
analysis are presented in Figure \ref{fig:regress}, which shows the marginal 
posterior probability density functions for the regression intercept ($A$) and 
slope ($B$) for each pre-MS model grid, along with the derived standard 
deviation of datapoints around the regression line (this scatter is assumed to 
be normally distributed with mean 0) and the associated correlation 
coefficient.  The last panel of Figure \ref{fig:regress} affirms the strong 
relationship determined from the correlation tests.  The 95\%\ confidence 
intervals for the regression lines are shown as shaded regions for the 
comparisons between $L_{\rm mm}$ and $M_{\ast}$ in Figure \ref{fig:mdvms}.

We find an intercept $A$=$-1.2\pm0.3$, $-1.8\pm0.2$, and $-1.6\pm0.2$ and slope 
$B$=$2.0\pm0.5$, $1.5\pm0.4$, and $1.7\pm0.4$ for the DM97, BCAH98, and SDF00 
pre-MS model grids, respectively (quoted uncertainties represent 95\%\ 
confidence intervals).  These fits suggest a typical 1.3\,mm flux density of 
$\sim$25\,mJy for 1\,$M_{\odot}$ stars, falling to only $\sim$4\,mJy at the 
peak of the host mass function ($\sim$0.3\,M$_{\odot}$).  The inferred slope 
($B$) values indicate that $L_{\rm mm}$ scales rather steeply with host mass 
(but see below).  That said, it should be obvious in Figure \ref{fig:mdvms} 
that there is a substantial scatter around the underlying relationship that is 
much larger than the conservative formal uncertainties on $L_{\rm mm}$ or 
$M_{\ast}$ can accommodate.  Assuming that this scatter is described by a 
normal distribution (with mean 0), the \citet{kelly07} regression method 
provides an inference of the variance of that distribution.  The corresponding 
standard deviation is $0.7\pm0.1$\,dex for all models: in short, at any given 
host mass, 68\%\ of the $L_{\rm mm}$ values span a factor of $\sim$5 on either 
side of the best-fit regression line.  

\begin{figure}[t!]
\epsscale{0.5}
\plotone{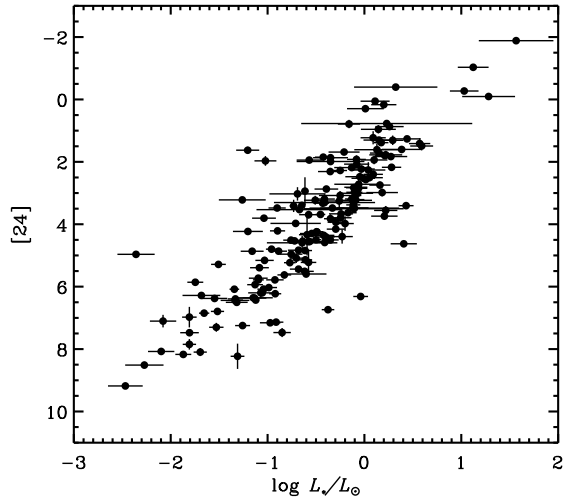}
\figcaption{Demonstration of the strong correlation between the amount of dust
emission in the mid-infrared \citep[shown here as the {\it Spitzer} 24\,$\mu$m 
apparent magnitude;][]{luhman10,rebull10}, which scales with the dust 
temperature $\langle T_d \rangle$, and the luminosity of the stellar host. 
\label{fig:templum}}
\end{figure}

A disk mass can be crudely estimated from a simple scaling of the mm-wave 
continuum luminosity \citep{beckwith90}.  Assuming the emitting dust is 
optically thin and isothermal,
\begin{equation}
\log{M_d} = \log{F_{\nu}} + 2\log{d} - \log{(\zeta \cdot \kappa_{\nu})} - \log{B_{\nu}(\langle T_d \rangle)},
\end{equation}
where $\kappa_{\nu}$ is the dust opacity, $\zeta$ is the dust-to-gas mass 
ratio, and $B_{\nu}(\langle T_d \rangle)$ is the Planck function at the average 
disk temperature.  For ease of comparison with other studies, we adopt the 
assumptions of \citet{aw05} and fix $d = 140$\,pc, $\zeta = 0.01$, and 
$\kappa_{\nu} = 2.3$\,cm$^2$\,g$^{-1}$ (at the reference wavelength of 
1.3\,mm).  The nominal conversion to $M_d$ advocated by \citet{aw05} assumes 
that $\langle T_d \rangle \approx 20$\,K is applicable for all disks.  In 
principle, the assumption of an average temperature in this calculation is well 
justified, since most of the mm-wave emission comes from cool material in the 
nearly isothermal outer disk \citep[radiative transfer calculations suggest a 
shallow $T_d$ profile, $\propto r^{-q}$ with $q \approx 
0.4$-0.6;][]{chiang97,dalessio98}.  However, there is good reason to expect 
that $\langle T_d \rangle$ increases with $M_{\ast}$.  In the regions where 
most of the mm-wave emission is generated, the disk is heated solely by 
irradiation from the central star.  Therefore, the local $\langle T_d \rangle$ 
is primarily set by $L_{\ast}$: all else being equal, disks around less 
luminous hosts should be cooler.  That hypothesis has a firm empirical backing, 
as demonstrated in Figure \ref{fig:templum} through the tight correlation of 
$L_{\ast}$ and the {\it Spitzer} 24\,$\mu$m emission: this mid-infrared 
emission is optically thick, and therefore roughly scales with $\langle T_d 
\rangle$.  To reflect that thermal relationship, we assume a reasonable scaling 
$\langle T_d \rangle \approx 25 (L_{\ast}/L_{\odot})^{1/4}$\,K (see Appendix B 
for $L_{\ast}$ estimates and Appendix C for a validation of this scaling using 
more sophisticated radiative transfer calculations).  Then, disk mass estimates 
(or upper limits) and uncertainties are determined from Equation 2 using the 
$F_{\rm mm}$ measurements in Table \ref{tab:Fmm}.  As with our calculations of 
\{$M_{\ast}$,$t_{\ast}$\}, we include an additional uncertainty on $M_d$ that 
accounts foran intrinsic ambiguity in the assumed distance of $\pm 20$\,pc 
($\sim$15\%).

\begin{figure}[t!]
\epsscale{1.0}
\plotone{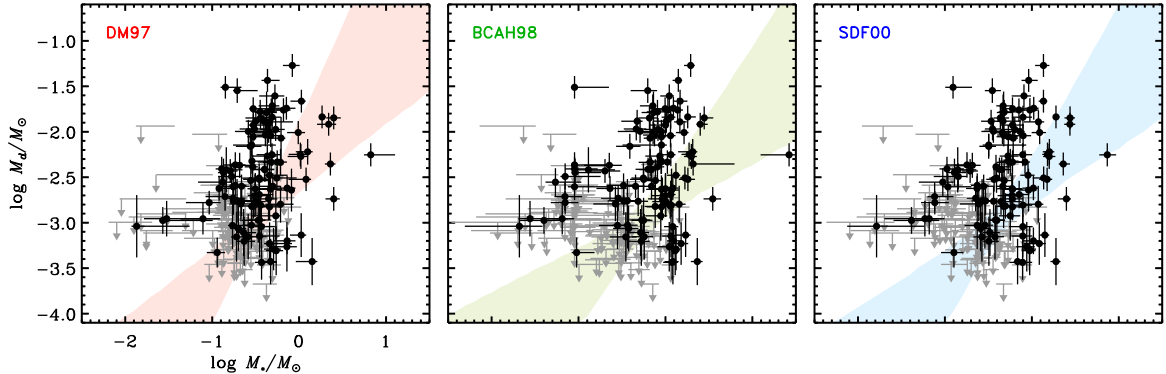}
\figcaption{As in Fig.~\ref{fig:mdvms}, but for $M_d$ and $M_{\ast}$.  
\label{fig:mdvms2}}
\end{figure}

\begin{figure}[t!]
\epsscale{1.0}
\plotone{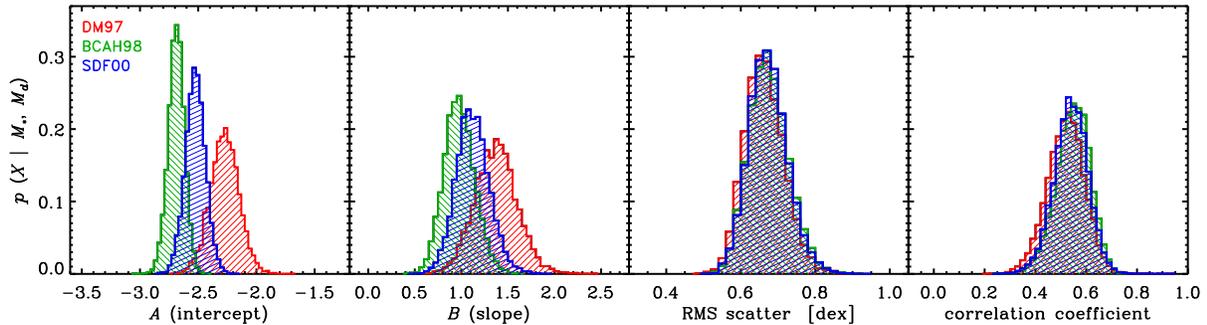}
\figcaption{As in Fig.~\ref{fig:regress}, but for $M_d$ and $M_{\ast}$.  
\label{fig:regress2}}
\end{figure}

\begin{figure}[t!]
\epsscale{0.5}
\plotone{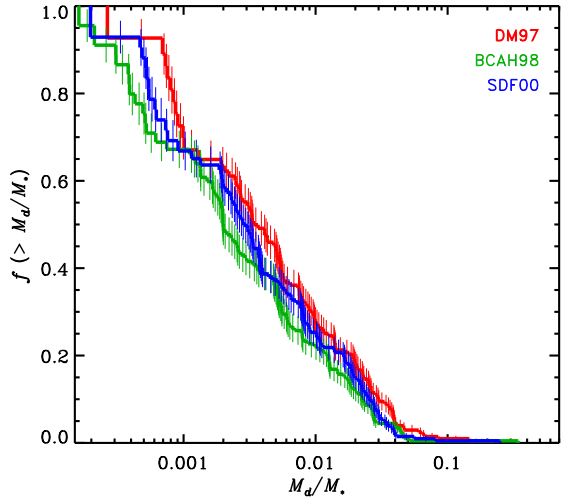}
\figcaption{The cumulative distribution of the disk-to-star mass ratios,
constructed from the Kaplan-Meier product-limit estimator to incorporate
censored datapoints.  The median mass ratio is $\sim$0.3\%, the upper
quartile of the distribution has $M_d/M_{\ast} \ge 1$\%: very few sources have
$M_d/M_{\ast} \ge 10$\%.  \label{fig:MdMs}}
\end{figure}

Figure \ref{fig:mdvms2} shows the direct comparisons of $M_d$ and $M_{\ast}$ 
for the full catalog sample for each of the three pre-MS model grids.  Again, 
we identify a strong correlation between these variables: the same censored 
statistical tests rule out the null hypothesis with similarly high confidence 
as for the $L_{\rm mm} \sim M_{\ast}$ relationship.  The results of a Bayesian 
linear regression on this correlation, where $\log{(M_d/M_{\odot})} = A + B 
\log{(M_{\ast}/M_{\odot})}$, are shown in Figure \ref{fig:regress2}; the shaded 
regions in Figure \ref{fig:mdvms2} represent the 95\%\ confidence intervals on 
the regression lines.  We find intercepts $A$=$-2.3\pm0.3$, $-2.7\pm0.2$, and 
$-2.5\pm0.2$ and slopes $B$=$1.4\pm0.5$, $1.0\pm0.4$, and $1.1\pm0.4$ for the 
DM97, BCAH98, and SDF00 model grids, respectively (quoted uncertainties are at 
the 95\%\ confidence level).  In short, this sample suggests a roughly linear 
scaling between disk and host star masses, $M_d \propto M_{\ast}$, with a 
typical disk-to-star mass ratio of $\sim$0.2--0.6\%.  For reference, the 
cumulative distributions of that mass ratio, $f$($>M_d/M_{\ast}$), for each 
pre-MS model grid are shown together in Figure \ref{fig:MdMs}.  The intrinsic 
scatter measured around this correlation is the same as was inferred from the 
relationship between $L_{\rm mm}$ and $M_{\ast}$, with a standard deviation of 
$0.7\pm0.1$\,dex (this corresponds to a full-width at half-maximum spread of a 
factor of $\sim$40 in the disk masses present at any given host mass).  

\subsubsection{Potential Sources of Scatter and Bias}

In the previous section, we demonstrated that there is a strong correlation 
between the measured mm-wave continuum luminosities from dust disks and the 
inferred masses of their stellar hosts.  It is natural to associate the shape 
of that correlation and the scatter around it with a corresponding relationship 
between $M_d$ and $M_{\ast}$ and an intrinsic distribution of $M_d$ at any 
given host mass (as was illustrated in 
Figs.~\ref{fig:mdvms}--\ref{fig:regress}).  However, the assumption of that 
scaling or those made in converting $L_{\rm mm}$ to $M_d$ could introduce 
scatter into the relationship, and might also bias the shape of the correlation 
derived from the regression analysis.  Here, we discuss five of the key issues 
that might influence these results: (1) a potentially more complex intrinsic 
relationship between $M_d$ and $M_{\ast}$; (2) additional dust temperature 
effects; (3) contamination from high optical depths; (4) assumptions about the 
emitting material (i.e., $\kappa_{\nu}$); and (5) variations in evolutionary 
state.  

The first issue is one of practical interpretation: we have implicitly assumed 
that the measured correlation between $L_{\rm mm}$ (or $M_d$) and $M_{\ast}$ 
can be functionally represented as a power-law scaling with a single index over 
the full range of host masses.  An intrinsic relationship that is different 
from the one imposed here would naturally bias our inferences of its shape and 
the associated scatter around it.  Although the current survey data do not 
necessarily warrant a more complex prescription to parameterize these 
relationships, they by no means rule out such a scenario either.  For example, 
bimodal or logistic functions might describe the correlations in Figures 
\ref{fig:mdvms} and \ref{fig:mdvms2} equally well, and could perhaps be 
characterized by less intrinsic scatter/outliers.  With the currently available 
data, the regression analysis is (rightfully) focused on the many 
low-$M_{\ast}$ targets that tend to have faint mm-wave emission, or, more 
commonly, upper limits on $L_{\rm mm}$ or $M_d$.  A more sensitive survey of 
this population that provides robust mm-wave continuum detections (or much 
deeper upper limits) should be able to discriminate between these (or other) 
alternatives.  

The nominal conversion to $M_d$ used above assumes a simple relationship
between the host luminosity and the average dust temperature is applicable for 
all disks in the sample.  That relationship implies that the slope of the 
correlation between $M_d$ and $M_{\ast}$ is substantially less steep than would 
be inferred from the normal assumption of a single $\langle T_d \rangle$ value
(which is identical to the $L_{\rm mm} \sim M_{\ast}$ relationship represented 
in Figs.~\ref{fig:mdvms} and \ref{fig:regress}).  Through this $\langle T_d 
\rangle$ correction, the intrinsic $L_{\ast}$ scatter in the H-R diagram (see 
Appendix B) contributes a dispersion of $\lesssim 0.2$\,dex in the \{$M_d$, 
$M_{\ast}$\}-plane, significantly less than what was measured in \S 3.2.2 
(0.7\,dex).  However, this $\langle T_d \rangle$ prescription only
considers the source of irradiation energy, and not the (possibly related)
efficiency at which the disk absorbs (thermalizes) it.  The latter is partly
set by the surface area of the disk that intercepts starlight, as determined by
the vertical distribution of small dust grains in the disk atmosphere.  If
those grains are coupled to a gas phase in hydrostatic equilibrium, the dust
height varies like $H \propto (\langle T_d \rangle/M_{\ast})^{0.5}$.  In this
sample, $L_{\ast} \propto M_{\ast}^{1.5-2}$, implying that $H$ is relatively
independent of $M_{\ast}$ (given the scaling between $\langle T_d \rangle$ and
$L_{\ast}$ advocated above).  In reality, the scaling between $H$, $\langle T_d 
\rangle$, and $M_{\ast}$ depends more intimately on the detailed coupling
between the gas and small grains.  Unfortunately, the processes that control
dust sedimentation and their impact on midplane temperatures are quite
complicated \citep[e.g.,][]{dullemond04,dalessio06}.  The diverse SED 
morphologies for disks (see Appendix A) suggest that dust settling probably 
represents an important contribution to the scatter in the relationship between 
$M_d$ and $M_{\ast}$ (of uncertain magnitude), but there is no indication that 
it should preferentially bias its slope.

The fundamental assumption in Eq.~(2) is that the mm-wave continuum emission is 
optically thin.  For a reasonable range of grain properties, this is a valid 
assumption for most disks \citep[e.g.,][]{beckwith90}.  Based on a fiducial 
model of disk structure, \citet{aw05} suggested that $\gtrsim$90\%\ of the 
total mm-wave luminosity is optically thin, although that fraction likely 
decreases for the brightest (most massive) sources.  Subsequent high angular 
resolution work has demonstrated that brighter disks tend to be more spatially 
extended, suggesting that most of their emission is generated from low column 
densities in the outer disk \citep{andrews10}.  Moreover, recent 
multi-wavelength observations of disks suggest that optical depths remain low 
even for the high column densities expected at small disk radii, due to the 
natural decrease in particle emissivities produced by dust grain growth 
\citep[e.g.,][]{guilloteau11,perez12}.  Given these resolved constraints, and 
since the under-estimate of $M_d$ induced by optically thick contamination is 
expected to be quite small in any case ($\lesssim 0.05$\,dex), it is clear that 
the assumption of optically thin emission made in Eq.~(2) has negligible 
influence on the shape of the correlation between $M_d$ and $M_{\ast}$ or the 
scatter around its corresponding best-fit regression lines.

The impact of the material properties of disks in the \{$M_d$, 
$M_{\ast}$\}-plane -- encapsulated here in the grain opacity, $\kappa_{\nu}$, 
and dust-to-gas mass ratio, $\zeta$ -- are considerably more difficult to 
predict generically.  Particle growth promotes a net decrease in the 
disk-averaged $\kappa_{\nu}$, while particle migration \citep[dominated by 
radial drift;][]{weidenschilling77} effectively decreases $\zeta$ over most of 
the disk volume \citep[e.g.,][]{takeuchi02,brauer08,birnstiel10}.  Models of 
this material evolution predict that growth timescales increase like $1/\zeta 
\sqrt{M_{\ast}}$ up to grain sizes ($a$) that are limited by fragmentation 
\citep[$a_{\rm max} \propto M_d$;][]{birnstiel11} or drift \citep[$a_{\rm max} 
\propto M_{\ast} M_d \, \zeta$;][]{birnstiel12}.  If all else is equal, the 
disks around more massive stellar hosts would have more top-heavy particle size 
distributions and (if $a_{\rm max} \gtrsim \lambda$) therefore lower 
$\kappa_{\nu}$ on average (this effect would be amplified by an underlying 
relationship between $M_d$ and $M_{\ast}$).  Considering our assumptions in 
Eq.~(2), this would imprint a negative bias on the regression slope inferred in 
\S 3.2.2: the true value of $B$ might be larger than suggested.  However, some 
caution in the generic interpretation of these models is warranted.  The 
quantitative effects of other parameters (including turbulence, $\zeta$, 
temperature, etc.) and how they mutually interact and scale with $M_{\ast}$ 
might dominate the model behavior.  In principle, the {\it sense} of any bias 
can be determined empirically by examining how the slope of the mm/radio-wave 
spectrum \citep[which is related to $\kappa_{\nu}$; e.g.,][]{beckwith91} 
depends on $M_{\ast}$.  In terms of the scatter in the relationship between 
$M_d$ and $M_{\ast}$, the natural dispersions in $\kappa_{\nu}$ and $\zeta$ 
would seem perfectly reasonable explanations.  For example, an RMS scatter of 
an order of magnitude in $a_{\rm max}$ and a factor of $\sim$2 in $\zeta$ can 
account for the measured dispersion.  

Aside from these effects of material evolution, the global viscous evolution of 
disk structures more generally might be expected to imprint some feature on the 
observed relationship between $M_d$ and $M_{\ast}$.  In the standard model for 
viscous evolution, disk masses decay at a rate $M_d \propto 
(t_{\ast}/t_s)^{-1/2}$, where $t_s$ is the viscous timescale 
\citep{hartmann98}.  In this scenario, a correlation between disk and host 
masses could be imposed at the formation epoch, or could be produced naturally 
if $t_s$ scales with $M_{\ast}$ and/or $t_{\ast}$ scales inversely with 
$M_{\ast}$.  There is no evidence for the latter; indeed, if anything, young 
clusters tend to exhibit preferentially older ages among their early-type 
members \citep[e.g., see][]{pecaut12}.  Formally, $t_s \propto \sqrt{M_{\ast}} 
/ \langle T_d \rangle$, which implies a negligible contribution to an intrinsic 
scaling between $M_d$ and $M_{\ast}$ (although a relationship between either 
$M_d$ or $M_{\ast}$ and the level of disk turbulence might imprint a feature).  
In essence, there is no obvious reason to assume that viscous evolution 
significantly biases the observed correlation.  Moreover, we find no evidence 
whatsoever for a relationship between $M_d$ and $t_{\ast}$ in this sample.  
Given the large uncertainties on the latter, this is not such a surprise: in \S 
3.2.1 we argued that there is no firm statistical evidence for an age gradient 
larger than $\pm$0.7\,dex in this sample.  Nevertheless, the permitted 
dispersion in $t_{\ast}$ is roughly sufficient to explain the measured scatter 
around the $M_d \propto M_{\ast}$ regression lines.

\section{Discussion}

We have significantly expanded the mm-wave photometry catalog of circumstellar 
disks in the Taurus star-forming region, using ``snapshot" continuum 
observations of $\sim$60 new sources with the SMA.  Folding these new disks, 
most of which have M-type stellar hosts, in with survey results in the 
literature, we have constructed a $\lambda = 1.3$\,mm luminosity census that is 
{\it complete} for the known Class II members in Taurus with spectral types 
earlier than M8.5 \citep[cf.,][]{luhman10}, down to a (3-$\sigma$) sensitivity 
limit of $\sim$3\,mJy (corresponding to a disk mass limit of $\sim$0.1 to 
1\,$M_{\rm Jup}$ for A- or M-type hosts, respectively, following the 
assumptions outlined in \S 3.2.2).  The derived luminosity distribution is 
found to be substantially different than in previous work, with a notable shift 
to weaker emission that was identified through a more robust understanding of 
individual disks in multiple systems \citep[facilitated by][]{harris12} and the 
incorporation of a large subset of previously un-observed faint disks that are 
preferentially hosted by low-mass stars and brown dwarfs.  Having remedied the 
selection bias of the original \citet{aw05} survey against these late-type 
members, we uncovered a strong correlation between the mm-wave luminosities 
emitted by dust disks and the spectral types (effective temperatures) of their 
hosts.

A Bayesian estimation technique developed by \citet{jorgensen05} has been used 
to convert the location of each individual host star in the H-R diagram into a 
corresponding mass and age, with reference to three representative grids of 
pre-MS evolution models \citep{dantona97,baraffe98,siess00}.  The results are 
used to demonstrate that the mm-wave dust continuum luminosity -- a proxy for 
the disk mass, through a simple scaling relation -- is strongly correlated with 
the mass (but not age) of its host, regardless of which pre-MS model grid is 
assumed.  A linear regression analysis in the (logarithmic) \{$M_d$, 
$M_{\ast}$\}-plane, conducted with the maximum likelihood estimator derived by 
\citet{kelly07} to properly account for measurement error, data censoring, and 
intrinsic scatter, favors a roughly linear scaling, $M_d \propto M_{\ast}$, 
with a typical disk-to-host mass ratio of $\sim$0.2--0.6\%.  Although the 
statistical evidence for this relationship is robust, its physical origins 
remain unclear: it may be a manifestation of the initial conditions imposed at 
disk formation, and/or a consequence of the combined effects from various 
evolutionary mechanisms that depend on the host mass.  Regardless, these 
results indicate that the host mass plays a fundamental factor in setting 
$M_d$, with an influence that is roughly equivalent to the presence of a close 
stellar companion \citep[e.g.,][]{jensen94,jensen96,harris12}.  

There is substantial scatter around the correlation between $M_d$ and 
$M_{\ast}$, which reflects the net effects of the intrinsic distributions of 
masses, temperatures, opacities, and evolutionary states in the sample.  
Assuming this dispersion is characterized by a normal distribution in the 
regression analysis, we measure a standard deviation of 0.7\,dex (a factor of 
$\sim$5 on either side of the scaling between $M_d$ and $M_{\ast}$).  For a 
relationship that spans only $\sim$3\,dex in each variable, this scatter is 
large in an absolute sense.  However, considering the substantial uncertainties 
we have regarding the physical processes that play important roles in setting 
$L_{\rm mm}$ -- particularly related to the material evolution of dust 
particles -- it is remarkable that the scatter is limited enough that a 
correlation can be identified at all.  These results suggest that the product 
of mm-wave dust opacities, disk masses, and temperatures has an intrinsic FWHM 
of a factor of $\sim$40 for any given $M_{\ast}$.  

Improved constraints on the basic morphology of the relationship between $M_d$
and $M_{\ast}$ will require a substantially more sensitive mm-wave photometry 
survey for M-type hosts.  For example, a census that probes 10$\times$ deeper 
than the data presented here (a relatively trivial time investment for ALMA) 
should be able to differentiate between the power-law scaling we have assumed 
and more dramatic alternatives (i.e., a precipitous drop in the mass scaling).  
Moreover, a multi-wavelength photometry census (preferably a long-wavelength 
complement to the catalog provided here) of the full Taurus sample is crucial, 
since it can provide fundamental constraints on the contribution of material 
evolution in these disks to the intrinsic scatter in the observed $M_d \propto 
M_{\ast}$ correlation.

\subsection{Implications for Planet Formation Models}

Although the analysis presented here represents the first statistically robust 
confirmation and characterization of a correlation between $M_d$ and 
$M_{\ast}$, it is by no means the first to suggest that such a relationship 
exists.  With a sample that was biased against hosts that span the lowest 
available $M_{\ast}$ decade and had a naturally sparse population at high 
$M_{\ast}$, \citet{aw05} found no evidence for the relationship inferred here 
(see their Fig.~9).  Some hints of a correlation have been tentatively noted 
when a larger (inhomogeneous) sample of Herbig AeBe stars is considered 
\citep[e.g.,][]{natta00,williams11}.\footnote{It is worth noting that both of 
these review articles ignored star-disk systems that did not have firm mm-wave 
{\it detections}, which was presumably part of the reason that they chose not 
to invest in a more involved correlation analysis or make any strong claims on 
the validity or characteristics of a relationship between $M_d$ and 
$M_{\ast}$.}  As others started to measure mm-wave luminosities from the disks 
around later type hosts, low detection rates and faint emission suggested a 
potential trend: \citet{scholz06} indicated that the disks around brown dwarf 
hosts are intrinsically low-mass \citep[see also][]{klein03}, and 
\citet{schaefer09} argued that hosts with spectral type later than M2 had 
systematically less mm-wave emission than their counterparts with earlier 
types.  However, there was a legitimate concern that the large population of 
M-type hosts that had {\it not} yet been observed might exhibit a different 
behavior than these limited subsamples.  

Despite the previous (and justified) reticence of the observational community,
an implicit linear scaling $M_d \propto M_{\ast}$ is pervasive in theoretical 
studies of planet formation.  The concept of such a scaling seems so natural in 
a gravitational sense that a physical basis for this foundational assumption is 
rarely offered in the literature.  Given the lingering uncertainties on the 
nature of disk formation and early evolution (and how those processes might 
depend on a changing $M_{\ast}$), there was good reason to question that 
assumption.  However, the results we have presented here now provide a firm 
empirical validation of the physical intuition behind it, thereby supporting 
some key conclusions inferred from planet formation theories that use disk 
properties as initial conditions.

In both the core accretion \citep{pollack96} and disk instability 
\citep{boss97} models, the overall efficiency of the giant planet formation 
process scales with the amount of mass available in the disk.  For the disk 
instability case, this is more of a threshold effect: regardless of the host 
mass, the potential for disk fragmentation and gaseous protoplanet formation is 
enhanced if $M_d/M_{\ast}$ is sufficiently large 
\citep[e.g.,][]{boss06,durisen07}.  However, our characterization of the $M_d 
\propto M_{\ast}$ scaling indicates that the disk-to-star mass ratio is 
basically independent of $M_{\ast}$.  Taking that as an initial condition, the 
disk instability model would predict a roughly constant gas giant planet 
fraction across the stellar mass distribution.\footnote{Note that the incidence 
of systems where disk instability is thought to operate, those with 
$M_d/M_{\ast} \gtrsim 0.1$, is very small in this sample: $<$5\%\ (see 
Fig.~\ref{fig:MdMs}).  Since $M_d$ estimates are likely to be under-estimated, 
the giant planet frequency expected from this model could reasonably be scaled 
up by a factor of a few (see \S 3.2.3 for more details).}  On the other hand, 
the efficiency of the core accretion model depends on a key timescale: a 
sufficiently massive solid core must be assembled before the gas disk 
dissipates.  This core growth is accelerated for disks with higher densities 
\citep[$\propto M_d$; e.g.,][]{ikoma00,hubickyj05,thommes08}, and sees modest 
benefits from shorter orbital periods ($\propto \sqrt{M_{\ast}}$ at a fixed 
semimajor axis) and larger potential formation zones \citep[$\sim M_{\ast}$; 
e.g.,][]{kennedy08}.  In this case, the intrinsic $M_d \propto M_{\ast}$ 
correlation found here implies that planet formation is inherently more likely 
around more massive stellar hosts 
\citep[e.g.,][]{laughlin04,ida05,kennedy08,alibert11}.  

The observational demographics of the exoplanet population strongly support 
this latter core accretion scenario: the incidence of giant planets orbiting 
$<$2.5\,AU from their hosts scales roughly linearly with $M_{\ast}$ 
\citep[e.g.,][]{johnson07,johnson10,bowler10}.  This relationship between stars 
and their planets not only fortifies the case for core accretion as the 
dominant pathway for giant planet formation, it also should be recognized as a 
clear manifestation of the initial association between the masses of young 
stars and their disks during the epoch of planet formation.

\subsection{Practical Consequences for Disk Evolution Studies}

From a more practical standpoint, we stressed in \S 1 that a proper accounting 
of selection biases is a fundamental requirement for comparative studies of 
disk evolution.  Previous work has suggested that disks in Orion 
\citep{mann10}, IC 348 \citep{lee11}, and Upper Sco \citep{mathews12} have 
systematically lower masses compared to Taurus \citep{aw05} and Ophiuchus 
\citep{aw07b}, presumably due to evolutionary effects related to their 
environment, particle growth, and disk dispersal, respectively.  However, these 
studies all relied on comparisons between incomplete samples with (usually 
unknown) selection biases.  Now that we have constructed a complete 
``reference" sample for Taurus disks and identified an important trend between 
the mm-wave luminosity ($\propto M_d$) and the stellar host type 
($\sim$$M_{\ast}$) that could introduce strong selection effects in the 
comparison samples, it is imperative to re-evaluate these claimed signatures of 
disk evolution in a more robust statistical framework.  

To properly compare the mm-wave luminosity distribution of an incomplete 
(potentially biased) sample from an arbitrary young cluster with the complete, 
reference sample derived here for Taurus, we must assume that both samples have 
the same intrinsic stellar mass function and multiplicity demographics.  Under 
those assumptions, a straightforward Monte Carlo simulation can be used to 
statistically compare the samples.  First, a trial reference sample of mm-wave 
luminosities is randomly drawn from the complete Taurus survey for a subset of 
sources that have the same distribution of host spectral types as the 
comparison sample.  In that random selection, multiplicity selection effects 
can be incorporated by treating close pairs as composite systems that are 
assigned the spectral type of the primary (as would be inferred with 
observations at any given resolution).  Next, the probability that these two 
subsamples are drawn from the same parent distribution, $p_{\emptyset}$, is 
evaluated using the standard suite of two-sample tests for censored datasets 
\citep[e.g.,][]{feigelson85}.  Then, that process is repeated for a large 
number ($\sim$10$^6$) of individual trials, and the results are used to 
construct a cumulative distribution of null hypothesis probabilities, 
$f(<p_{\emptyset})$.  

\begin{figure}[t!]
\epsscale{0.5}
\plotone{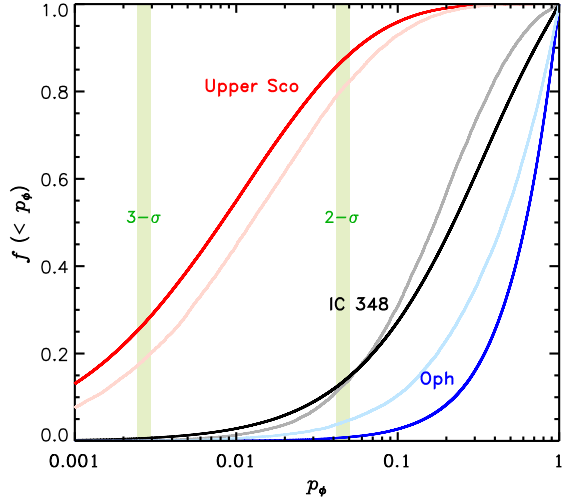}
\figcaption{The cumulative distributions of null hypothesis probabilities, 
$p_{\emptyset}$ -- the probability that an incomplete comparison sample of 
mm-wave luminosities is drawn from the same parent distribution as the Taurus 
reference sample -- constructed from two-sample tests for censored datasets in 
Monte Carlo simulations designed to account for selection biases related to 
multiplicity and host classification.  The lighter curves include the effects 
of $L_{\rm mm}$ uncertainties, using random draws from a normal distribution 
with mean $L_{\rm mm}$ and variance $\sigma_L^2$ for individual sources in each 
sample.  For reference, we mark the nominal 2-$\sigma$ and 3-$\sigma$ 
probabilities that the two samples are different.  \label{fig:comparisons}}
\end{figure}

This technique was applied to the incomplete comparison samples of the Class II 
disks catalogued in Ophiuchus \citep{aw07b}, IC 348 \citep{lee11,espaillat12}, 
and Upper Sco \citep{mathews12}, with reference to the Taurus sample provided 
here (in computing luminosities, we adopt distances of 125, 350, and 145\,pc, 
respectively, for the comparison samples).\footnote{The Orion sample compiled 
by \citet{mann10} lacks a sufficient number of hosts with firm spectral 
classifications to make a meaningful comparison.  Nevertheless, the key 
environmental effect noted in that work -- a systematic decrease in mm-wave 
luminosities for hosts near OB stars -- is independent of host properties.}  
The resulting cumulative distribution functions, $f(<p_{\emptyset})$, 
constructed from the Peto-Prentice test are shown in Figure 
\ref{fig:comparisons} (the Gehan and logrank tests give similar results).  
Vertical green bars denote the $p_{\emptyset} = 0.0455$ and 0.0027 levels, 
which correspond to the nominal ``2-$\sigma$" and ``3-$\sigma$" probabilities, 
respectively, that the two samples being compared are different. 

The cumulative distributions in Figure \ref{fig:comparisons} indicate that the 
incomplete Ophiuchus and IC 348 samples have mm-wave luminosity distributions 
that are statistically indistinguishable from the Taurus reference sample 
compiled here, while the Upper Sco sample appears to have marginally 
($\sim$2.5-$\sigma$) different (in this case lower) luminosities on average.  
For Ophiuchus and Upper Sco, these results are in good agreement with the 
original analyses of \citet{aw07b} and \citet{mathews12} -- although the latter 
case does not quite represent as striking a difference as originally 
suggested, primarily due to the limited size (only 20 Class II disks) of the 
comparison sample.  In contrast, \citet{lee11} claimed that the IC 348 disks 
have a mm-wave luminosity distribution that is significantly shifted to values 
$\sim$20$\times$ lower than its Taurus counterparts.  Those results are not 
borne out by our analysis, which only finds a $<$2-$\sigma$ difference between 
the samples in $>$85\%\ of the Monte Carlo trials and suggests a much smaller 
$L_{\rm mm}$ shift between the two samples (only $\sim$30\%\ on average).  This 
discrepancy is due solely to selection effects in the IC 348 sample: most of 
the \citet{lee11} disk targets have mid/late M-type hosts, and we have 
demonstrated that such targets are systematically weaker mm-wave sources in the 
Taurus reference sample.  When this host selection bias is taken into 
account, there is no available evidence for evolution in the mm-wave luminosity 
distribution between the Taurus and IC 348 samples.

\section{Summary}

We have used a ``snapshot" $\lambda = 1.3$\,mm survey with the SMA to obtain 
mm-wave photometry measurements of the continuum emission from protoplanetary 
dust disks, primarily around M-type hosts, for 60 previously un-observed 
targets in the Taurus star-forming region.  By combining these new results with 
previous measurements in the literature, we constructed a mm-wave continuum 
luminosity census for the Taurus region that is statistically complete for 
Class II disks for hosts with spectral types earlier than M8.5, with a 
(3-$\sigma$) depth of approximately 3\,mJy.  This $L_{\rm mm}$ catalog was then 
used to explore the potential mass relationship between dust disks and their 
stellar hosts.  The key conclusions that were drawn from the analysis of these 
data include: 

\begin{itemize}

\item There is a strong correlation between the mm-wave luminosities from dust 
disks and the spectral types (or effective temperatures) of their stellar 
hosts.  Employing the \citet{jorgensen05} method for estimating stellar masses 
and ages from the behavior of pre-main sequence model grids in the H-R diagram, 
we found that this correlation corresponds to $L_{\rm mm} \propto 
M_{\ast}^{1.5-2.0}$, with a typical 1.3\,mm flux density of 
$\sim$25\,$(d/140{\rm pc})^{-2}$\,mJy for a 1\,$M_{\odot}$ stellar host.  The 
steepness of that relationship should be an important consideration in the 
planning for future mm-wave continuum surveys of disks in other young clusters.

\item Assuming a reasonable scaling of the dust temperature with host 
luminosity (where $\langle T_d \rangle \propto L_{\ast}^{1/4}$), we associated 
the correlation between $L_{\rm mm}$ and spectral type with an intrinsic, 
roughly linear, scaling between the disk and host masses, $M_d \propto 
M_{\ast}$.  The typical disk-to-star mass ratio is $\sim$0.2--0.6\%.  There is 
a large dispersion around this correlation -- $\sim$0.7\,dex, corresponding to 
a FWHM range of a factor of $\sim$40 in $M_d$ at any given $M_{\ast}$ -- 
contributed by the inherent diversity in temperatures, dust opacities, and 
evolutionary states in the Taurus sample. 

\item After considering the predictions from planet formation models, we 
suggested that the linear correlation between $M_d$ and $M_{\ast}$ determined 
here likely represents the origin of the correlation between the giant planet 
frequency and host mass that has been identified in the exoplanet population.  
This fundamental demographic association between stellar hosts and both the 
initial conditions (disks) and final outcomes (exoplanets) of the planet 
formation process provides strong, albeit circumstantial, support for the 
theoretical ``core accretion" model.  

\item Finally, we urged caution in the comparative analysis of incomplete, 
potentially biased, mm-wave luminosity samples with the goal of placing 
constraints on disk evolution mechanisms.  We used a set of Monte Carlo 
simulations with two-sample tests designed for censored datasets to demonstrate 
that a selection bias {\it toward} late-type hosts, and not dust evolution, is 
most likely responsible for the perceived difference in $f(L_{\rm mm})$ between 
the IC 348 and Taurus Class II disk populations.  In the future, such 
comparisons should rely on this kind of statistical analysis when comparing 
potentially biased sub-samples, while striving to assemble complete samples 
that ideally have well-characterized selection effects.

\end{itemize}

\acknowledgments We are very grateful to Rahul Shetty and Brandon Kelly for 
their invaluable advice on a Bayesian approach to linear regression analysis, 
to Jonathan Williams for providing the full set of mm-wave measurements for IC 
348 members, and to Til Birnstiel and Ana{\"e}lle Maury for helpful comments 
and critiques.  We also would like to thank the referee for a prompt and 
constructive review that helped fortify the presentation of our key 
conclusions.  ALK was supported by a Clay Fellowship from the 
Harvard-Smithsonian Center for Astrophysics.  The SMA is a joint project 
between the Smithsonian Astrophysical Observatory and the Academia Sinica 
Institute of Astronomy and Astrophysics and is funded by the Smithsonian 
Institution and the Academia Sinica.  The research involved in this article has 
made extensive use of the SIMBAD database, operated at CDS, Strasbourg, France 
and the NASA/IPAC Infrared Science Archive, which is operated by the Jet 
Propulsion Laboratory, California Institute of Technology, under contract with 
the National Aeronautics and Space Administration.

\appendix

\section{Spectral Energy Distributions}

In our efforts to determine some key observable stellar parameters (see 
Appendix B) and place the mm-wave measurements presented here in an appropriate 
context, we assembled a broadband reference SED for each system in this sample, 
using measurements available in the literature.  Those SEDs are made available 
for each individual source in a simple electronic ASCII format (an associated 
{\tt readme} file explains the columns in each table), and are displayed 
together in Figure 13.  

Optical photometry measurements were collated from monitoring surveys 
\citep{bastian79,rydgren84,vrba86,vrba89,vrba93,walker87,bouvier88,bouvier93,bouvier95,herbst94,herbst99,petrov99,oudmaijer01,grankin07,grankin08}, smaller 
compilations for individual targets \citep{rydgren83,myers87,gregorio-hetem92,hartigan94,strom94,kenyon95,torres95,duchene99,white01,dewinter01,briceno02,vieira03,beskrovnaya04,luhman04,kraus06,guieu06,audard07,herczeg08,luhman09b,duchene10}, as well as the Sloan Digital Sky Survey 
\citep[SDSS--8;][]{adelman-mccarthy11} and the Carlsberg Merdian Catalog 
\citep[CMC14;][]{evans02}.  All $UBVRI$ or 
$u^{\prime}g^{\prime}r^{\prime}i^{\prime}z^{\prime}$ measurements were 
converted to the Johnson-Cousins \citep{bessell79} or SDSS \citep{fukugita96} 
photometric systems, respectively.  Near-infrared data was compiled primarily 
from the Two Micron All-Sky Survey point source catalog 
\citep[2MASS;][]{skrutskie06}, and supplemented with other data as appropriate 
\citep{kenyon94,kenyon95,malfait98,eiroa01,eiroa02,woitas01,white01,kraus07,konopacky07,duchene10,mccabe11,dahm11,schaefer12}.  All $JHK$ measurements were 
transformed to the 2MASS--$JHK_s$ photometric system \citep{carpenter01a}.  

\begin{figure}
\epsscale{1.0}
\figurenum{13a}
\plotone{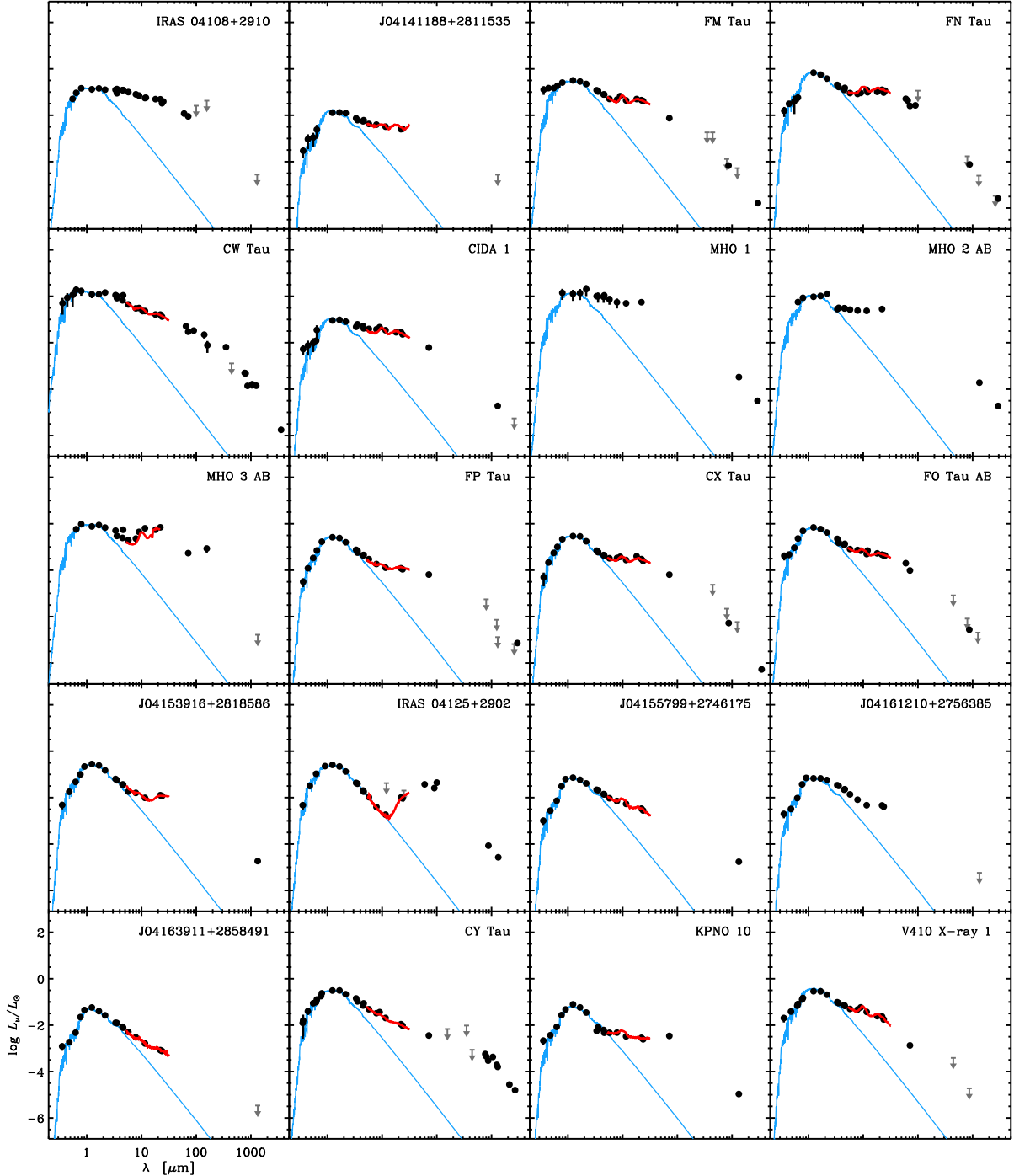}
\figcaption{Spectral energy distributions for the 179 Class II systems in 
Taurus that have mm-wave photometry measurements (note the ordinate axis is 
$L_{\nu} = 4\pi d^2 \nu F_{\nu}$ in $L_{\odot}$ units).  Each SED has been 
de-reddened based on the best-fit $A_V$ values listed below in Table 
\ref{tab:HR}.  Upper limits are shown as grey arrows (at the 3-$\sigma$ 
level).  Blue curves represent the best-fit stellar photosphere model (or 
composite for multiple systems); see Appendix B for details on the modeling 
process.  Red curves mark the {\it Spitzer} IRS spectra that are available from 
the Infrared Science Archive.  \label{fig:SEDs}}
\end{figure}

\begin{figure}
\epsscale{1.0}
\figurenum{13b}
\plotone{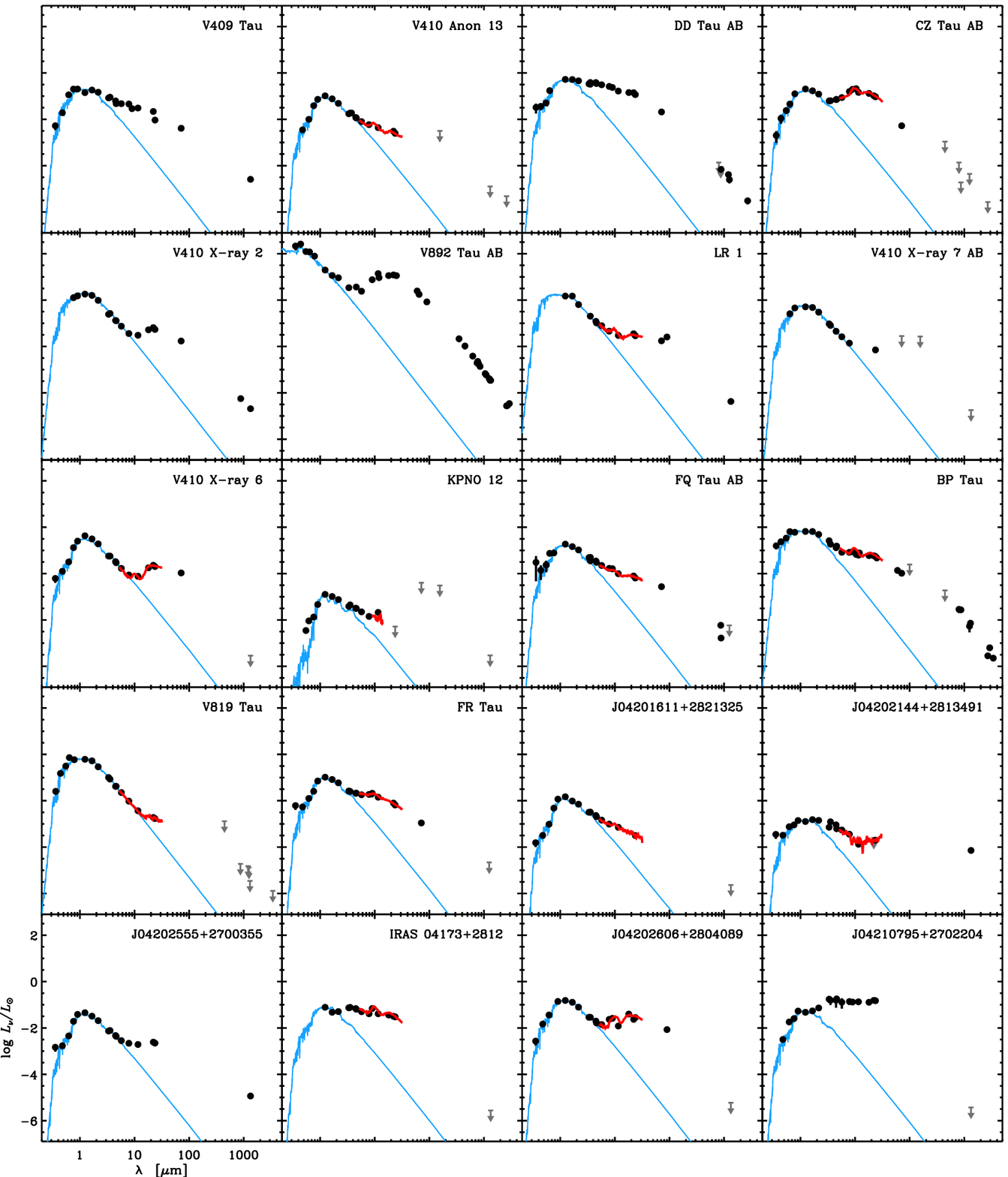}
\figcaption{As in panel (a).}
\end{figure}

\begin{figure}
\epsscale{1.0}
\figurenum{13c}
\plotone{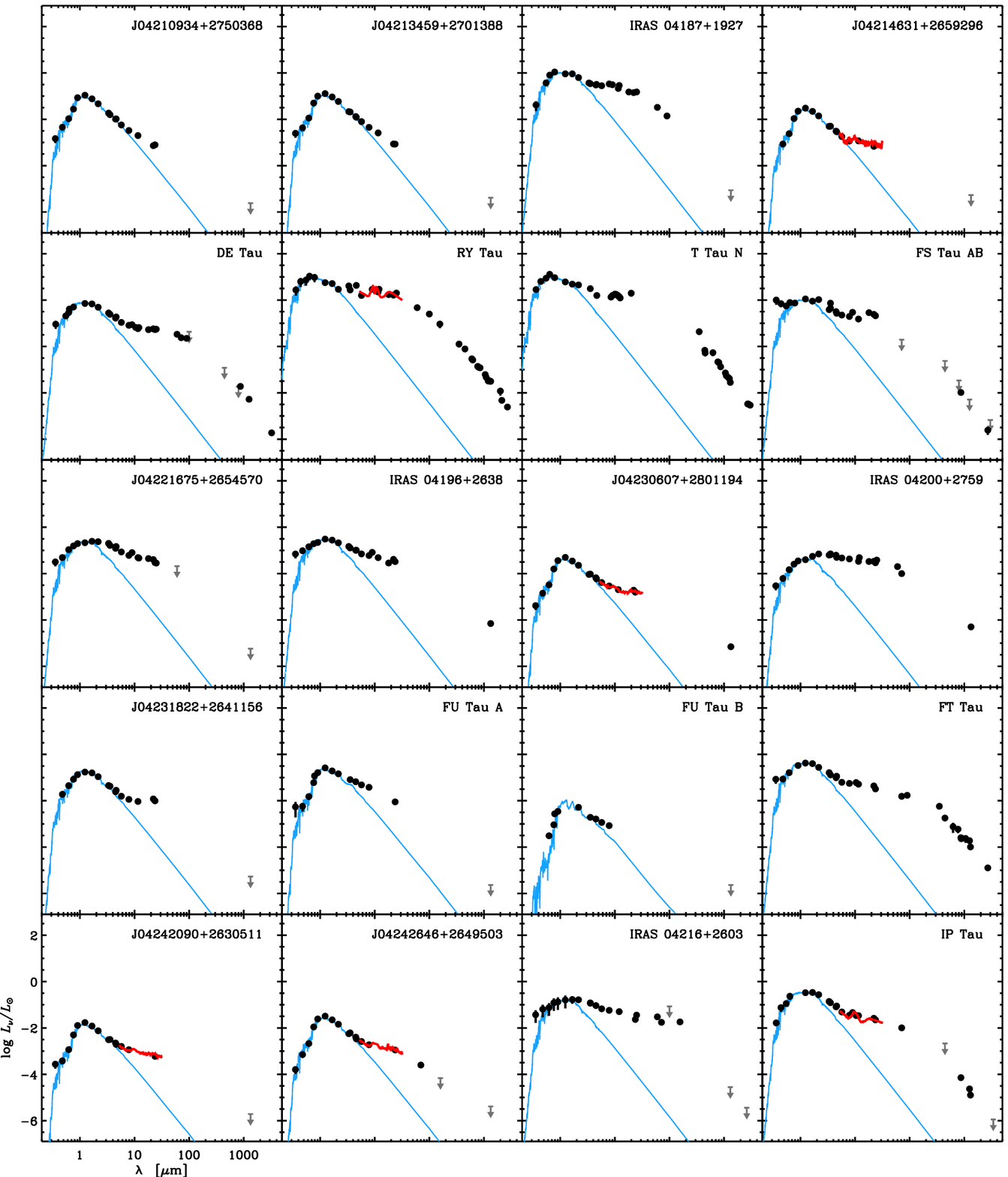}
\figcaption{As in panel (a).}
\end{figure}

\begin{figure}
\epsscale{1.0}
\figurenum{13d}
\plotone{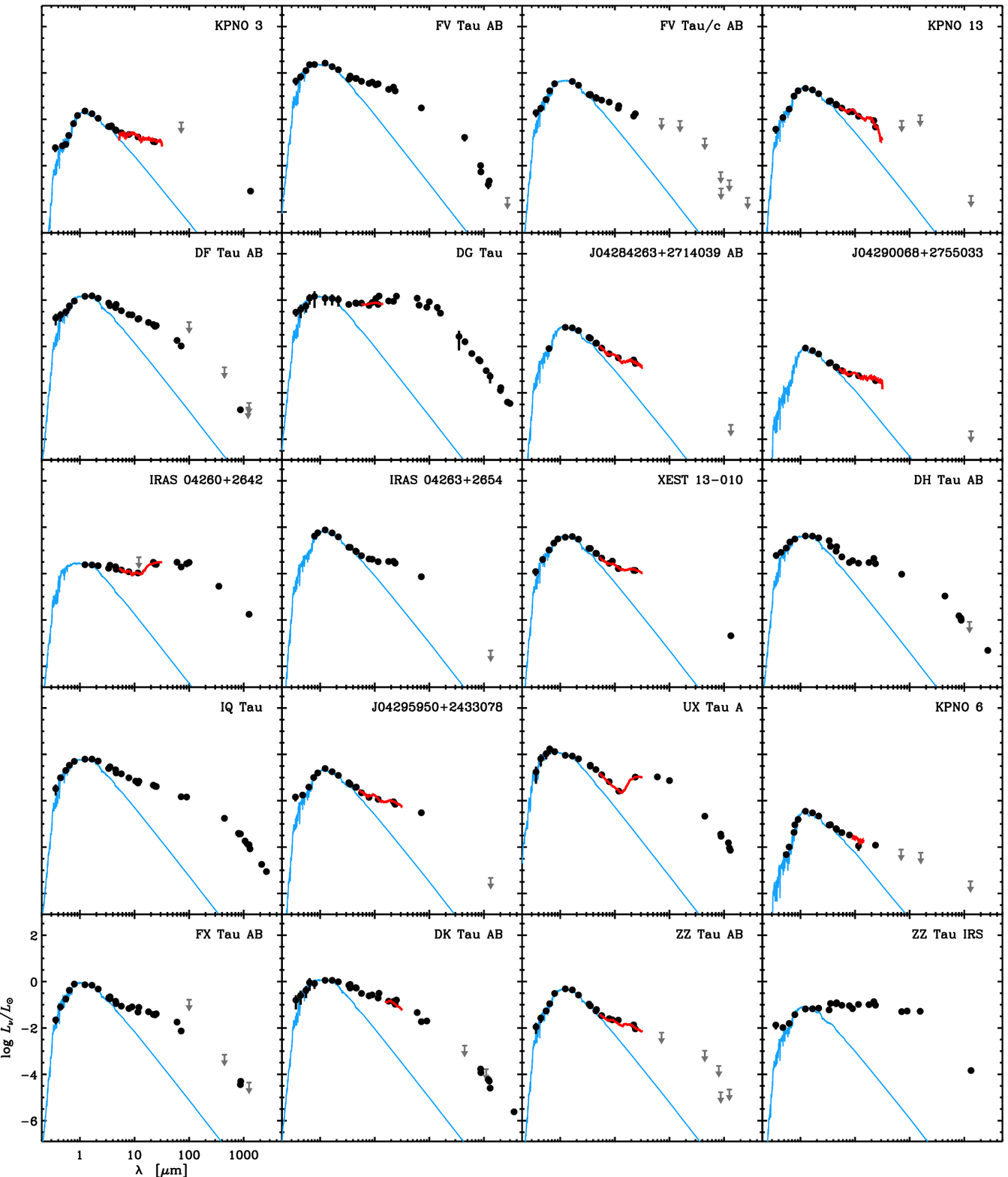}
\figcaption{As in panel (a).}
\end{figure}

\begin{figure}
\epsscale{1.0}
\figurenum{13e}
\plotone{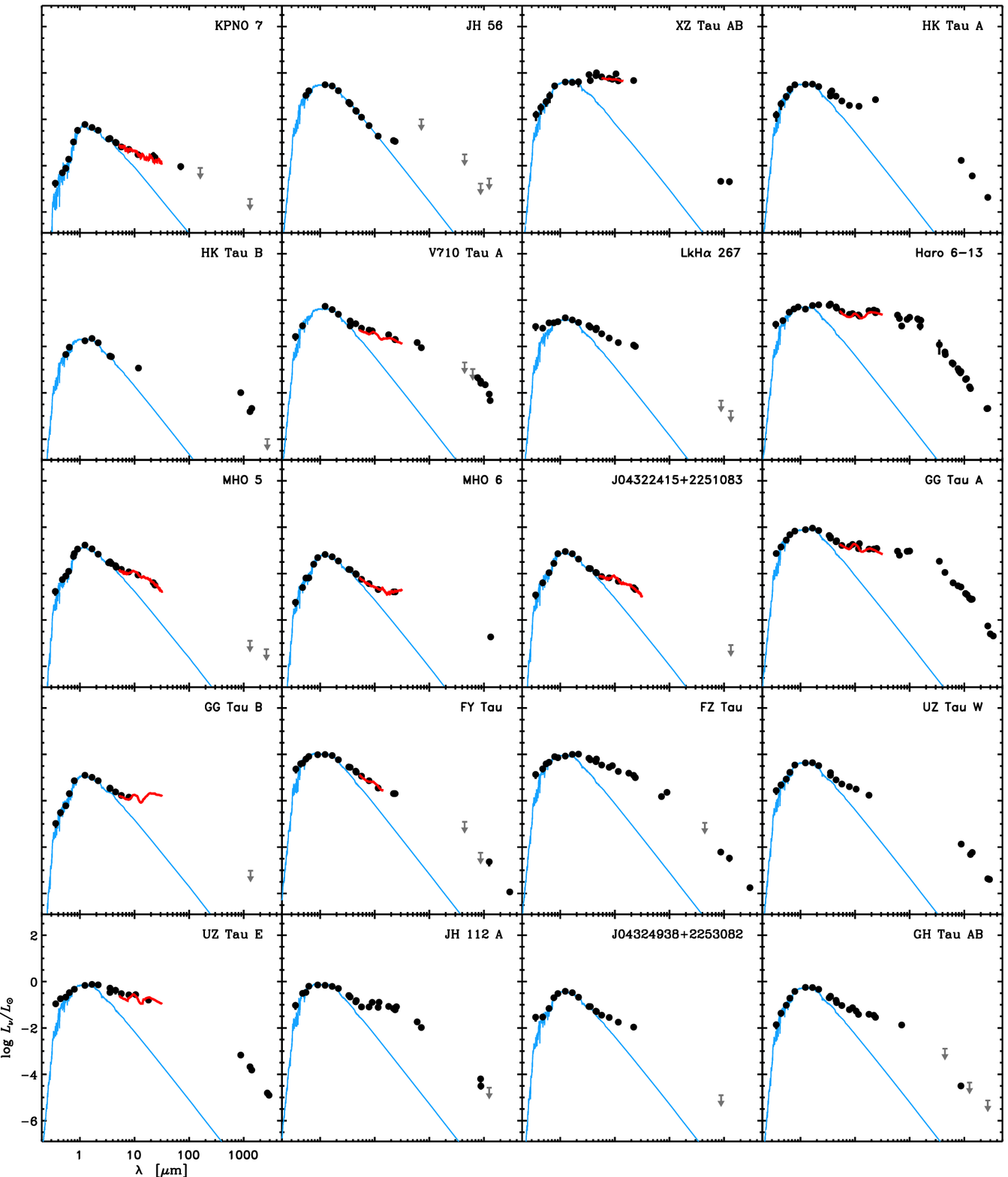}
\figcaption{As in panel (a).  Note that the panel for V710 Tau A utilizes some 
composite, unresolved measurements for the A--B pair, but a stellar photosphere 
model for V710 Tau B has been subtracted (see Table \ref{tab:HR} for V710 Tau 
B stellar parameters).}
\end{figure}

\begin{figure}
\epsscale{1.0}
\figurenum{13f}
\plotone{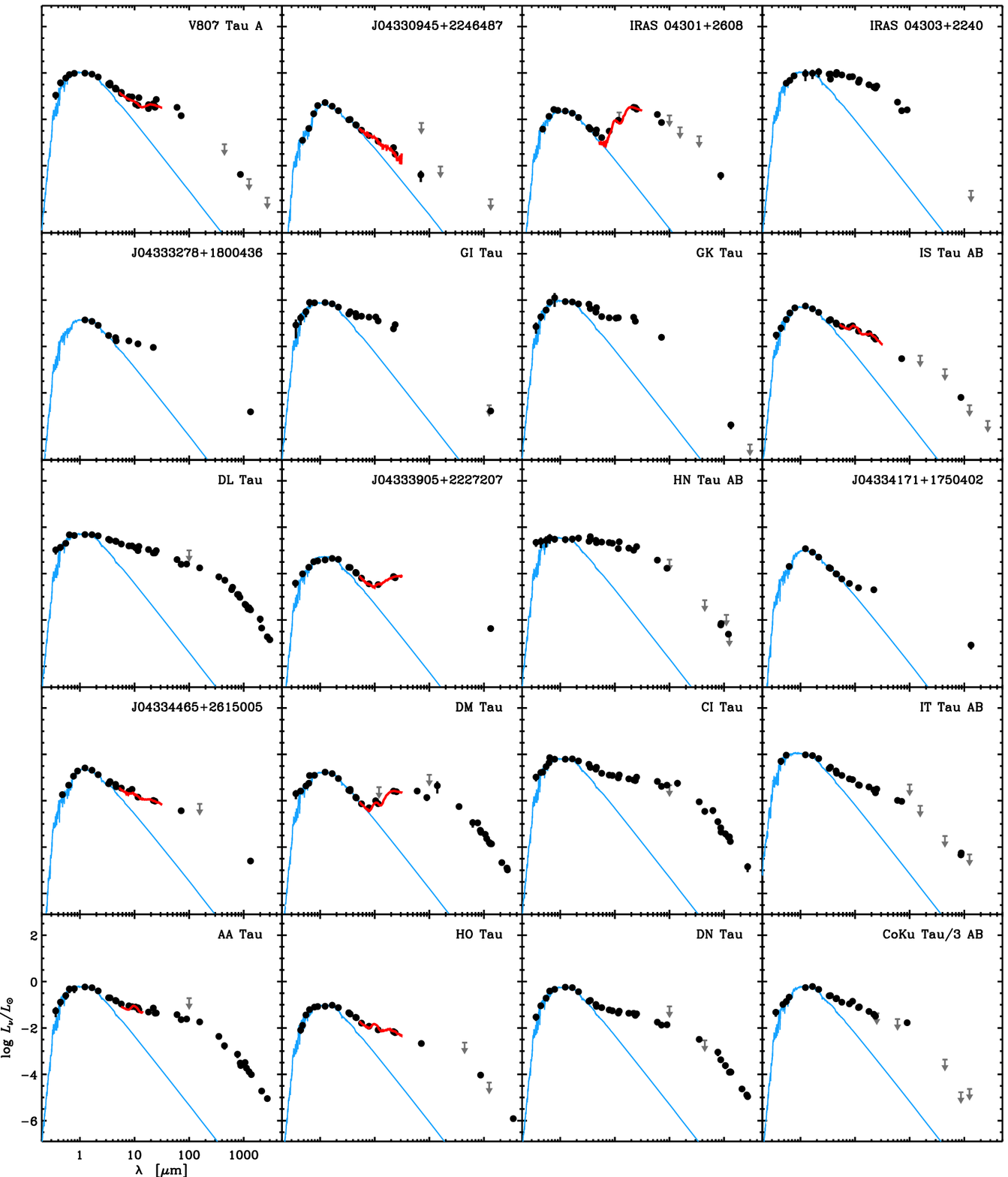}
\figcaption{As in panel (a).  Note that the panel for V807 Tau A utilizes some
composite, unresolved measurements for the A--B system, but a stellar 
photosphere model for V807 Tau B has been subtracted (see Table \ref{tab:HR} 
for V807 Tau B stellar parameters).}
\end{figure}

\begin{figure}
\epsscale{1.0}
\figurenum{13g}
\plotone{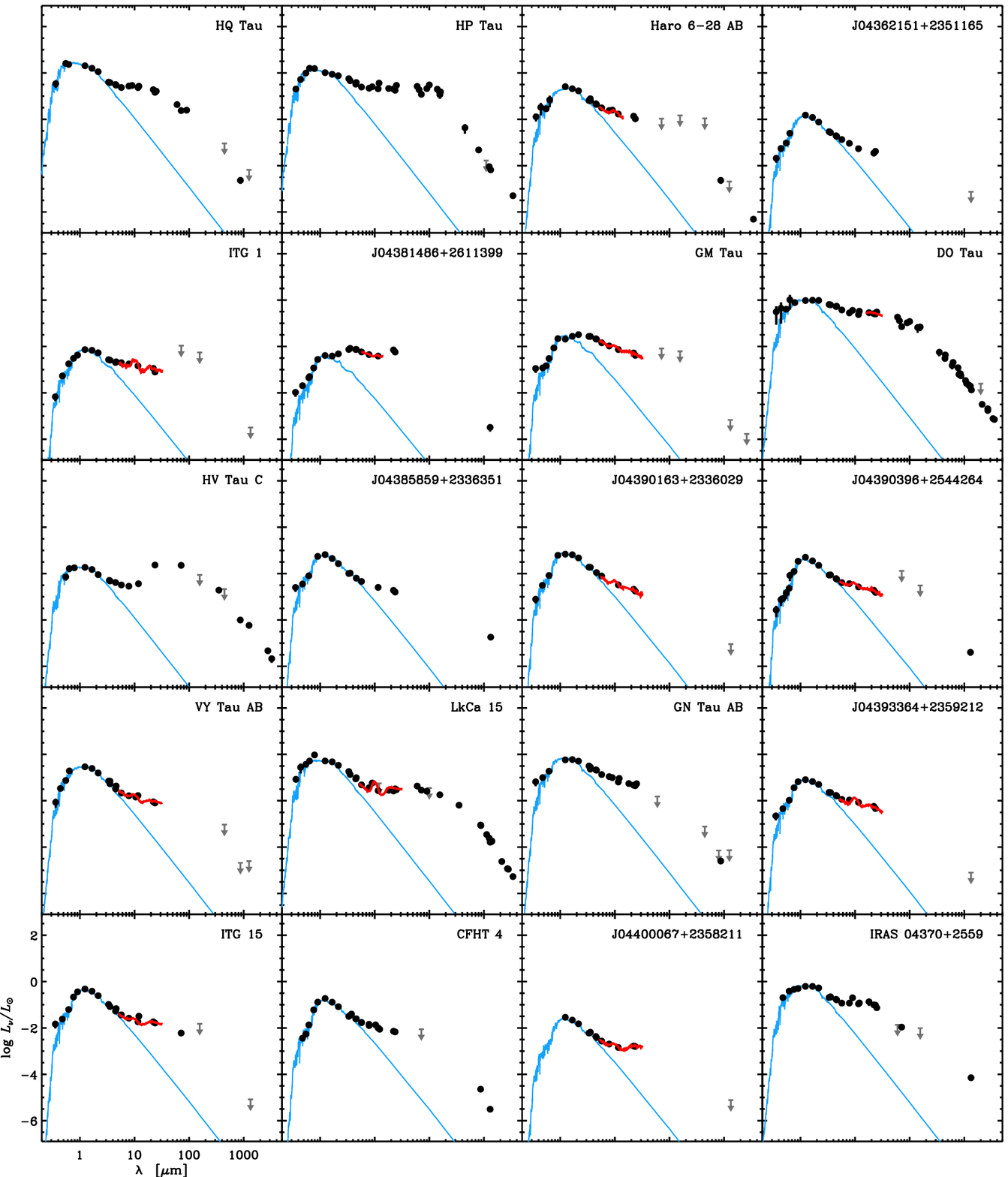}
\figcaption{As in panel (a).}
\end{figure}

\begin{figure}
\epsscale{1.0}
\figurenum{13h}
\plotone{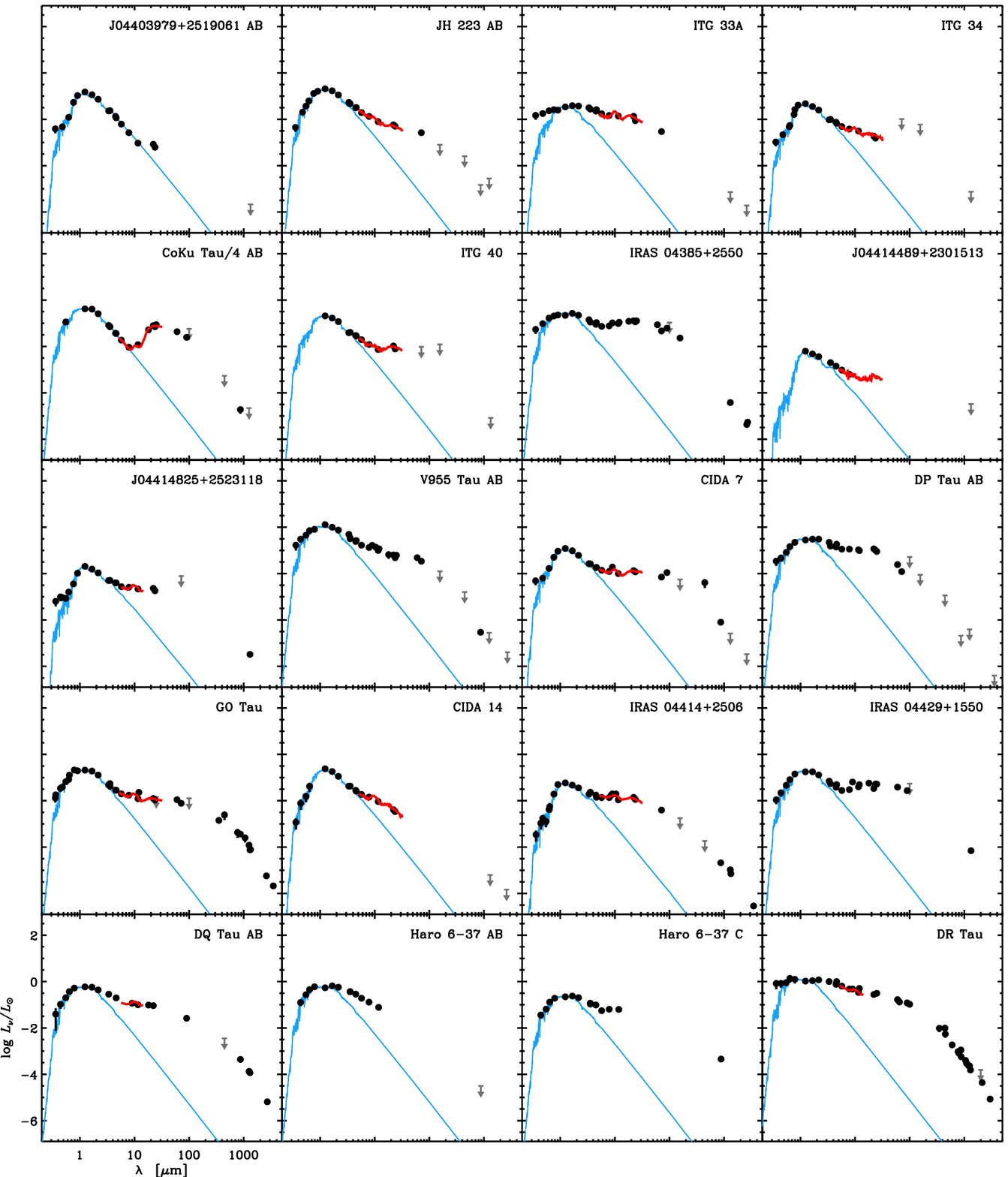}
\figcaption{As in panel (a).}
\end{figure}

\begin{figure}
\epsscale{1.0}
\figurenum{13i}
\plotone{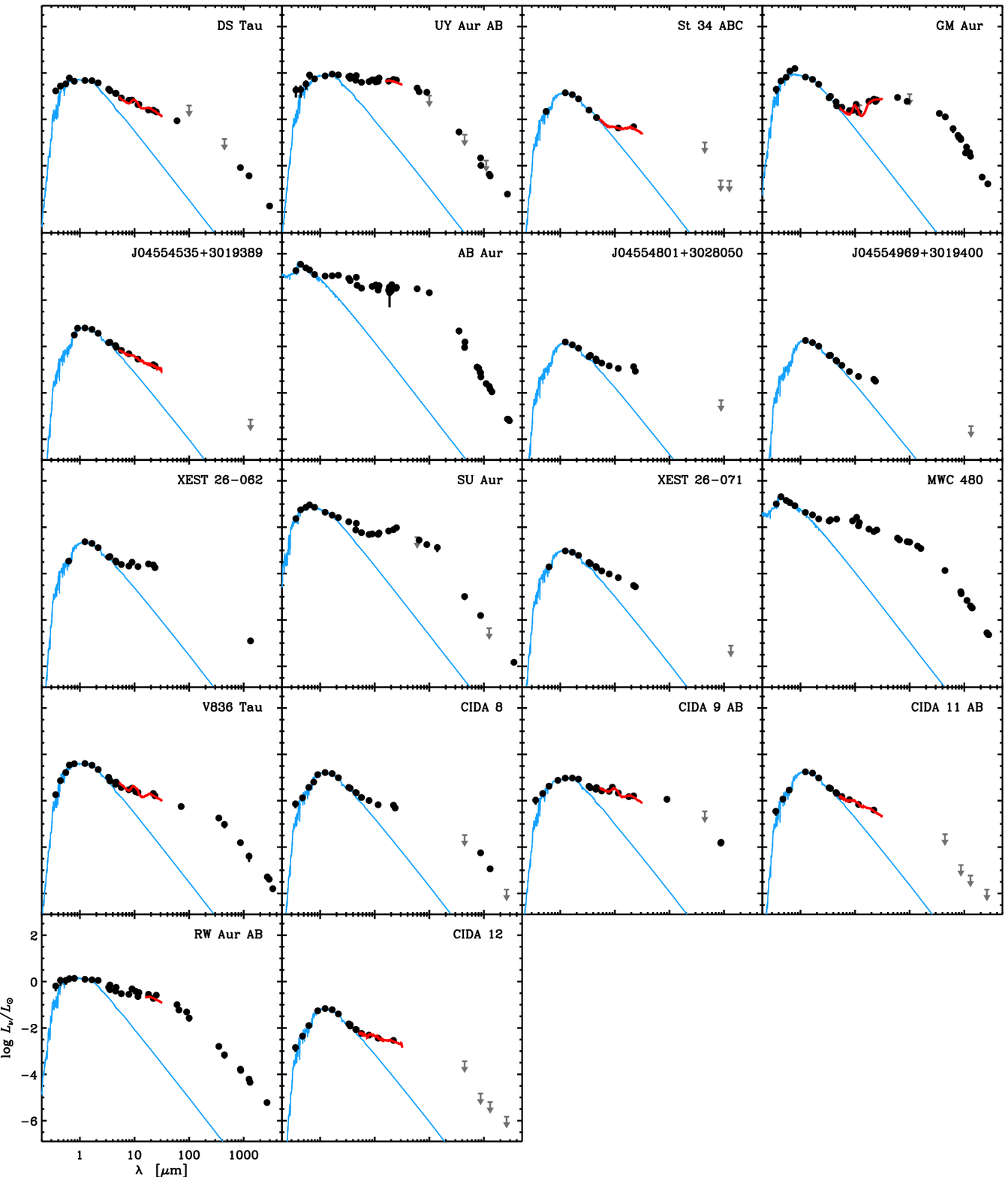}
\figcaption{As in panel (a).}
\end{figure}

Each datapoint in the optical/near-infrared SED is associated with a 
measurement error in the flux density (or magnitude), $\sigma_m$, and a 
systematic calibration error, $\sigma_s \approx f \cdot F_{\nu}$ (where $f$ 
usually reflects the fractional uncertainty in the filter zero-point, magnitude 
system conversions, etc.; for the wavelengths of interest here, $f$ is small, 
$\sim$2-5\%).  However, T Tauri stars are also known (indeed partially {\it 
defined}) to be variable, which naturally increases the scatter in the SED to 
levels that are typically larger than these formal observational 
uncertainties.  Unfortunately, simultaneous observations at a range of 
wavelengths sufficient to robustly determine the stellar parameters of interest 
at each epoch are practically non-existent.  Nevertheless, the scatter induced 
by variability can be accomodated by treating it as an additional error term, 
even if we are ignorant of its origins or cadence.  Of the Taurus Class II 
members in this sample, $\sim$40 have a large number of independent optical 
measurements that we use to construct a composite distribution of magnitudes 
($m$) at each wavelength, using an average of a sequence of normal 
distributions,
\begin{equation}
\langle \mathcal{F}(m) \rangle = \frac{1}{N} \sum_{i=1}^N \frac{1}{\sqrt{2\pi}\sigma_i} \exp \left[- \frac{(m - \mu_i)^2}{\sigma_i^2} \right],
\end{equation}
where $N$ is the number of measurements, $\mu_i$ are the reported magnitudes,
and $\sigma_i^2 = \sigma_{m,i}^2 + \sigma_{s,i}^2$.  In most cases, $\langle 
\mathcal{F}(m) \rangle$ has a single, clearly defined (if asymmetric) peak that
we then define as the representative magnitude in the ensemble, $\langle m 
\rangle$.  The cumulative distribution of $\langle \mathcal{F}(m) \rangle$ is 
used to assign a noise term caused by variability, $\sigma_v$, that represents 
the central 68\%\ of the composite distribution.  In essence, this is a 
slightly more sophisticated version of a weighted mean and standard deviation.
For the many other sources that do not have sufficient data available to 
measure $\langle \mathcal{F}(m) \rangle$, we assign a mean $\sigma_v$ term 
determined from this subset: $\bar{\sigma}_v = 0.33$\,mags in $B$, 0.25\,mags 
in $V$ and $R$, and 0.11\,mags in $I$.\footnote{For reference, we also found 
$\bar{\sigma}_v \approx 0.4$\,mags in $U$, but the $\langle \mathcal{F}(m) 
\rangle$ were much more complex than at longer wavelengths (due to accretion 
variability).  There is circumstantial evidence that the scatter in $R$ is 
partially tied to variability in H$\alpha$, in that the sources with larger 
scatter tend to have higher H$\alpha$ equivalent widths (or luminosities).}  
Near-infrared variability studies are rare for Taurus sources, so we adopt the 
results of a similar analysis of analogous sources in Orion by 
\citet{carpenter01}, where $\bar{\sigma}_v \approx 0.09$, 0.08, and 0.07\,mags 
at $J$, $H$, and $K_s$ (in the 2MASS system).  Flux densities are calculated 
assuming the zero-points advocated by \citet{bessell79} -- $F_U = 1810$, $F_B = 
4260$, $F_V = 3640$, $F_R = 3080$, and $F_I = 2550$\,Jy -- and in the 2MASS 
point source catalog explanatory supplement \citep{cutri03} -- $F_J = 1594$, 
$F_H = 1024$, and $F_{K_s} = 667$\,Jy.  

At longer infrared wavelengths, the SEDs shown in Figure \ref{fig:SEDs} were 
collected primarily from {\it Spitzer} imaging surveys 
\citep{luhman10,rebull10}, the {\it Wide-field Infrared Survey Explorer} ({\it 
WISE}) all-sky catalog \citep{wright10}, the {\it AKARI} point source catalogs 
\citep{ishihari10}, and the {\it Infrared Astronomical Satellite} ({\it IRAS}) 
point source catalog \citep{beichman88}, with some additional ground- and 
space-based measurements \citep{strom89,hillenbrand92,malfait98,white01,liu03,jayawardhana03,chen03,apai04,metchev04,pantin05,mccabe06,mccabe11,kundurthy06,marinas06,bouy08,monnier08,ratzka09,duchene10,honda10,skemer10,wahhaj10,grafe11,harvey12}.  In addition to the new data presented here (see Table \ref{tab:obs}), 
integrated flux density measurements in the submillimeter--radio spectrum are 
compiled from various catalogs in the literature \citep{weintraub89,adams90,beckwith90,beckwith91,altenhoff94,jewitt94,mannings94,jensen94,koerner95,osterloh95,ohashi96,momose96,momose10,kitamura96,kitamura02,dutrey96,dutrey03,mannings97,hogerheijde97,difrancesco97,henning98,dent98,akeson98,guilloteau98,guilloteau99,guilloteau11,looney00,duvert00,motte01,klein03,jensen03,duchene03,duchene10,corder05,pietu05,pietu06,pietu11,lin06,moriarty-schieven06,scholz06,hamidouche06,aw05,aw07a,rodmann06,cabrit06,bouy08,hughes08,hughes09,schaefer09,isella09,isella10,isella12,hamidouche10,oberg10,ricci10,ricci12,ricci13,andrews11,phan-bao11,sandell11,cieza12,harris12,harris13}.

\section{Effective Temperatures and Stellar Luminosities}

\begin{figure}[t!]
\epsscale{1.1}
\figurenum{14}
\plottwo{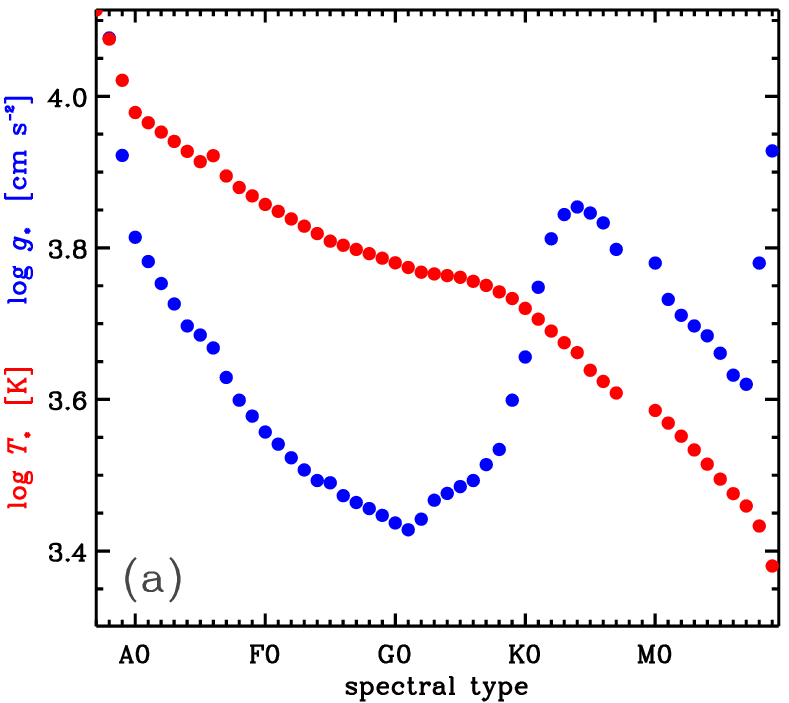}{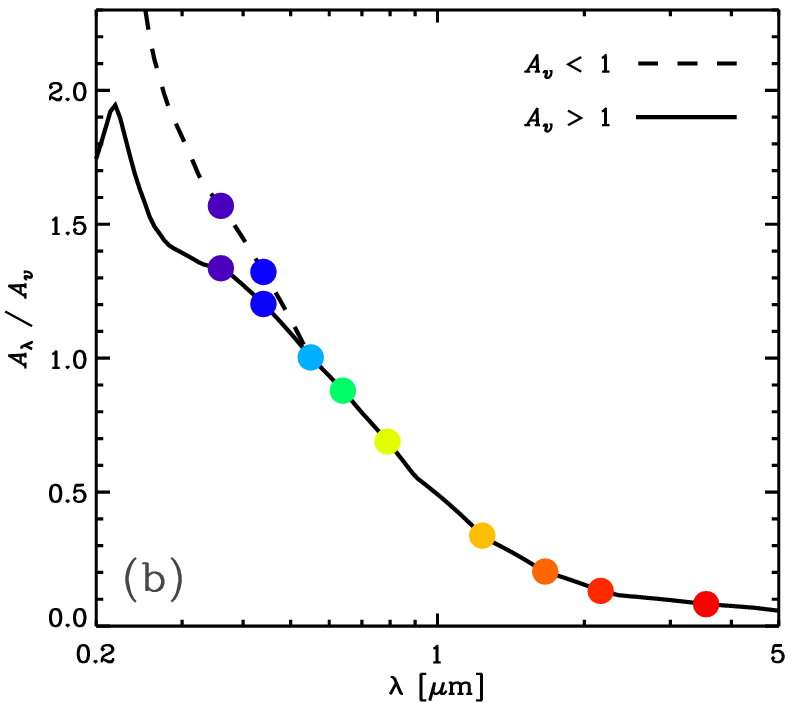}
\figcaption{(a) The assumed correspondances between $T_{\ast}$ ({\it red}) or
$\log{\,g_{\ast}}$ ({\it blue}) and spectral type used in our derivation of the observable quantities of the H-R diagram.  (b) The extinction curves used in
our derivations of stellar luminosities, based on the work by \citet{mathis90} 
and \citet{mcclure09}.  Colored dots represent the common optical/near-infrared
filter bands, for reference.  \label{fig:starmethods}}
\end{figure}

As mentioned in \S 3.1, effective temperatures are assigned based on the 
correspondances with spectral classification advocated by 
\citet{schmidt-kaler82}, \citet{straizys92}, and \citet{luhman99}, as shown in 
Figure \ref{fig:starmethods}(a) ({\it red} curve).  In most cases, the spectral 
types in Table \ref{tab:Fmm} were culled from the \citet{luhman10} catalog 
(alternative references are in the table notes).  We assume the default 
uncertainty on a type is $\pm$1 sub-class, and associate a normal (1\,$\sigma$) 
uncertainty on $T_{\ast}$ based on the (largest) temperature difference 
corresponding to that range of types.  Some early-type stars have a wider range 
of published spectral types (see Table \ref{tab:Fmm}), which are appropriately 
reflected in their adopted $T_{\ast}$ uncertainties.  The published 
classifications for many mid/late-M stars are more accurate than assumed here 
\citep[$\pm$0.5 sub-classes or better; e.g.,][]{luhman09}, but we feel our 
conservative approach is warranted due to the lingering uncertainty in the type 
-- temperature conversion.  Spectral classifications for individual stars in 17 
close multiple systems are not available \citep[for others, see][or the Table 
\ref{tab:Fmm} notes]{duchene99,hartigan03}.  In those cases, we assign the 
composite type to the primary and constrain the type/temperature of the 
secondary with two limiting assumptions: (1) $T_{\ast,1} \ge T_{\ast,2}$ (by 
definition), and (2) the ratio of stellar radii is $\ge$1, so $T_{\ast,2} \ge 
T_{\ast,1} (L_{\ast,2}/L_{\ast,1})^{1/4}$.  The mean optical/near-infrared 
contrast is used as a proxy for the luminosity ratio in condition (2), and then 
a temperature and uncertainty are defined as the midpoint and half the range of 
permissible $T_{\ast,2}$ values, respectively (corresponding spectral types are 
assigned from the conversion in Fig.~\ref{fig:starmethods}a).  This approach 
works well for binaries where individual types are known, and still properly 
reflects the intrinsic classification uncertainty.  It is similar to the method 
used by \citet{kraus11}, but has the benefit of not referencing pre-MS 
evolution models.  Finally, three sources (FT Tau, ITG 1, and IRAS 04370+2559) 
remain unclassified: types and uncertainties were assigned based on the rough 
accounting bins of \citet{luhman10}. 

We chose to not rely on previous assignments of luminosities because of the 
heterogenous measurement techniques adopted in the literature.  Instead, we 
developed a method that estimates luminosities ($L_{\ast}$) and extinctions 
($A_V$) -- and their uncertainties -- by fitting spectral templates of stellar 
photosphere models to the optical and near-infrared SED \citep[similar in 
spirit to the methods used by][]{bailer-jones11,brown11,dario12}.  The basic 
approach is simple.  First, we select a template spectrum based on the spectral 
type, using the conversions in Figure \ref{fig:starmethods}(a) to an 
appropriate $T_{\ast}$ and $\log{\,g_{\ast}}$.  The gravity dependence of the 
spectrum makes little difference in terms of the broadband photometry being 
fitted: deviations from these values by up to 0.3\,dex produce only small 
\{$L_{\ast}$, $A_V$\} changes that are well within the uncertainties determined 
here.  The $\log{\,g_{\ast}}$ behavior in Figure \ref{fig:starmethods}(a) ({\it 
blue} curve) corresponds to the mean behavior of the three pre-MS evolution 
models utilized in \S 3.2.1 at an age of $\sim$2\,Myr.  We adopt the {\tt 
NextGen}/{\tt PHOENIX}-based \citep{hauschildt99} ``BT--settl" model templates 
\citep{allard03,allard11} assuming solar metallicity, although other models 
\citep[e.g.,][]{lejeune97} produce comparable end results.  Next, we scale and 
redden the template for a given \{$L_{\ast}$, $A_V$\}.  The adopted 
extinction curves are shown in Figure \ref{fig:starmethods}(b).  At low 
extinctions ($A_V < 1$) we use the standard \citet{cardelli89} curve ($R_V = 
3.1$).  At higher $A_V$ we prefer the \citet{mcclure09} relation, which is 
based on {\it Spitzer} observations through dark clouds \citep[and equivalent 
at the wavelengths of interest to the $R_V = 5$ case of][]{mathis90}.  Then, we 
generate a synthetic model SED by convolving the scaled, reddened template with 
the relevant set of filter profiles to facilitate a proper comparison with the 
data.

\begin{figure}[t!]
\epsscale{1.0}
\figurenum{15}
\plotone{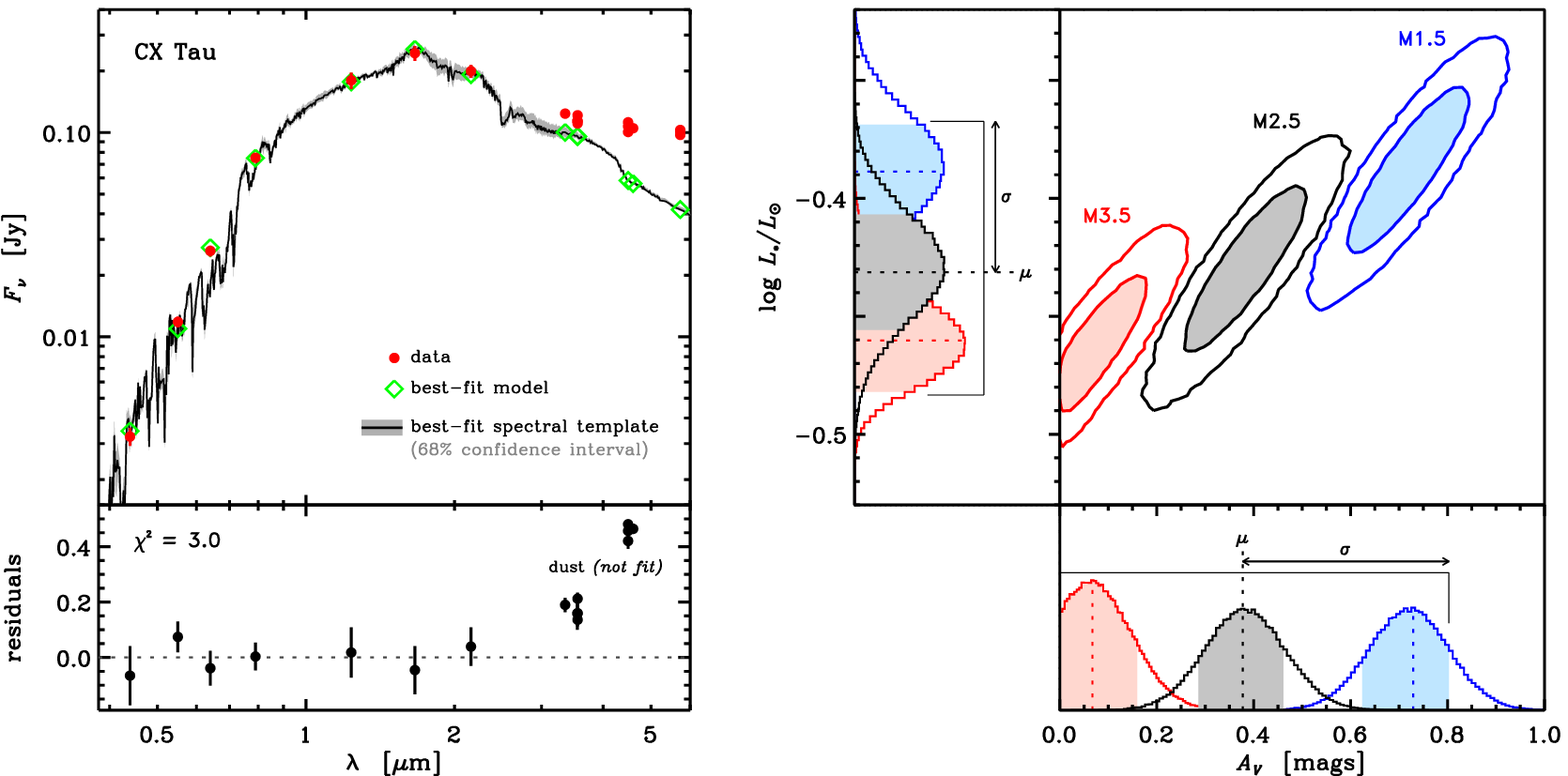}
\figcaption{A graphical illustration of the fitting technique used to derive
\{$\log{\,L_{\ast}}$, $A_v$\} (e.g., for CX Tau, spectral type M2.5$\pm$1).
({\it left}) The broadband SED for CX Tau ({\it red} points) are compared with
the best-fit scaled, reddened {\tt NextGen} spectral template ({\it black}
curve: {\it gray} shading represents the variations allowed within the 68\%\
confidence intervals of \{$T_{\ast}$, $\log{\,L_{\ast}}$, $A_v$\}), and the
best-fit SED model ({\it green} points) determined by convolving the template
with the appropriate observational filter profiles.  The fractional residuals
of that model fit are shown in the lower panel.  ({\it right}) The marginal
distributions $p(\log{\,L_{\ast}}|\{F_{\nu}\})$ (ordinate),
$p(A_V|\{F_{\nu}\})$ (abscissa), and their joint two-dimensional projections
derived from the MCMC fitting for the nominal spectral type ({\it gray}) and
the extrema classifications of $\pm$1 sub-class ({\it blue} and {\it red},
respectively).  Contours are drawn at 68 and 95\%\ confidence intervals; the
68\%\ level is shaded.  The diagrams mark the adopted best-fit value for each
parameter ($\mu$) and its associated uncertainty ($\sigma$), as described in
the text.  \label{fig:starexample}}
\end{figure}

Each model is evaluated with a likelihood function, $\ln{(\mathcal{L})} \propto 
-\chi^2/2$, that is directly related to the standard $\chi^2$ statistic summed 
over a range of wavelengths (0.4--4\,$\mu$m), depending on availability and a 
visual inspection for dust contamination in the infrared.\footnote{Note that 
$U$ or $u^{\prime}$ data are never used, due to contributions from accretion 
energy \citep[e.g.,][]{gullbring98}.}  A Monte Carlo Markov Chain (MCMC) 
technique is employed to maximize $\mathcal{L}$ and determine the marginal 
posterior probability density functions $p(\log{\,L_{\ast}}|\{F_{\nu}\})$ and 
$p(A_V|\{F_{\nu}\})$, where \{$F_{\nu}$\} represents the set of fitted flux 
density measurements, using the \citet{goodman10} ensemble sampler as 
implemented by \citet{foreman-mackey13}.  However, this parameter estimation 
process does not directly account for $T_{\ast}$ uncertainties on the 
inferences of \{$\log{\,L_{\ast}}$, $A_v$\}.  In principle that could be 
achieved by treating $T_{\ast}$ as another free parameter, but the broadband 
SED is a poor effective temperature diagnostic compared to a detailed spectrum 
(from which a spectral type, and therefore $T_{\ast}$, is derived here).  
Instead, we fold in the $T_{\ast}$ uncertainty with a discretized, brute-force 
approach: the entire modeling process is simply repeated two additional times, 
for templates with effective temperatures $T_{\ast} \pm \sigma(T_{\ast})$.  The 
best-fit \{$\log{\,L_{\ast}}$, $A_v$\} are identified from the peaks of their 
marginal posterior distributions for the nominal $T_{\ast}$, and their 
``1-$\sigma$" uncertainties are assigned from the (maximal) distance to the 
68\%\ confidence interval in the posterior distributions for the 
$T_{\ast}\pm\sigma(T_{\ast})$ results: mathematically, $\sigma(x) = {\rm max} | 
\{ [\mu(x) \pm \sigma(x)]_{T_{\ast} \pm \sigma(T_{\ast})} - [\mu(x)]_{T_{\ast}} 
\} |$, where $x \in \{\log{\,L_{\ast}}, A_v\}$, $\mu(x)$ denotes the best-fit 
$x$, and $\sigma(x)$ the 68\%\ uncertainty derived from $p(x|\{F_{\nu}\})$.  An 
example of the fitting process is summarized graphically in Figure 
\ref{fig:starexample}, for the arbitrary case of CX Tau.  

\begin{figure}[t!]
\epsscale{0.5}
\figurenum{16}
\plotone{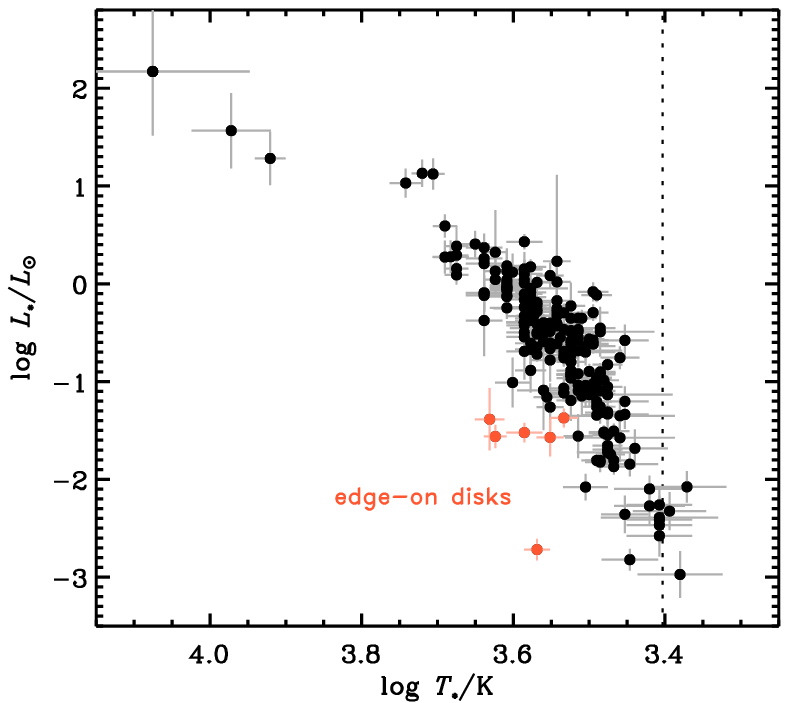}
\figcaption{The H-R diagram for Class II sources in Taurus, as compiled in
Table \ref{tab:stars}.  The dotted vertical line marks the completeness limit
in spectral type for this catalog.  \label{fig:HR}}
\end{figure}

We should point out that the adopted modeling procedure explicitly ignores any 
excess luminosity from accretion.  High spectral resolution measurements from 
the ultraviolet through the near-infrared suggest that a veiling continuum can 
contribute a non-negligible amount to the observed emission, particularly at 
blue-optical wavelengths 
\citep[e.g.,][]{basri90,hartigan91,hartigan95,gullbring98}.  Recently, 
\citet{fischer11} demonstrated that even the near-infrared peak of the typical 
SED can exhibit an excess \citep[for high accretion rates; see also]{cieza05}.  
In some pathological cases, that excess can dominate the $L_{\ast}$ values 
determined with our adopted procedure (and certainly affects the context in 
which the $A_V$ values should be considered), but more often the accretion 
contribution plays a lesser role.  Regardless, the $L_{\ast}$ values inferred 
here are likely slight over-estimates of the true stellar luminosities.  
Therefore, the stellar ages estimated in \S 3.2 would be correspondingly 
younger than the true values.  However, we do not expect accretion to 
significantly bias the derived $M_{\ast}$ values that are derived with the 
results of these stellar model fits, since the pre-MS model mass tracks scale 
roughly with $T_{\ast}$ (e.g., see Fig.~\ref{fig:HRhowto}).  

Following the prescription outlined above, we derived estimates of the 
fundamental stellar ``observables" \{$\log{\,T_{\ast}}$, $\log{\,L_{\ast}}$, 
$A_V$\} and their uncertainties for the full sample catalog: the results are 
compiled in Table \ref{tab:HR} and shown together in the H-R diagram in Figure 
\ref{fig:HR}.  Special care was taken in the fitting for stars in close 
multiple systems.  When component-resolved photometry at {\it more than two 
wavelengths} was not available \citep[e.g., not covered by][or similar 
work]{white01,woitas01}, we assumed that both components had the same 
extinction and fit both the contrasts and the composite SED simultaneously to 
estimate individual luminosities (see notes in Table \ref{tab:stars}).  There 
are 6 stars in this sample that are either known or strongly suspected to have 
edge-on disk orientations, either from direct high-resolution imaging 
\citep[HK Tau B, HV Tau C;][]{duchene03,mccabe11} or a combination of an 
anomalously low $L_{\ast}$ (a $>$2-$\sigma$ deviation from the mean luminosity 
at that spectral type), high $A_V$, or unusual optical/near-infrared SED 
morphology (J04202144+2813491, IRAS 04260+2642, IRAS 04301+2608, and ITG 
33A).\footnote{Note that LkH$\alpha$ 267 exhibits an anomalously low extinction 
in the optical bands that are inconsistent with the observed luminosity in the 
near-infrared \citep{kraus09}.  We elect to fit the $i^{\prime}z^{\prime}JH$ 
data only here, but note that this source might also be consistent with an 
edge-on disk (or perhaps some remnant envelope material).}  Those sources are 
marked in Figure \ref{fig:HR} and Table \ref{tab:HR}.  The inferred $L_{\ast}$ 
values for these cases are not representative of the true stellar luminosities, 
due to substantial obscuration from disk material along the line-of-sight to 
the central stars.  When estimating stellar masses and ages for these sources 
(see \S 3.2), we assign a luminosity and associated uncertainty based on the 
weighted mean and standard deviation of the $L_{\ast}$ values for all other 
sources within $\pm$1 spectral type sub-class.

\section{Comments on the Dust Temperature Scaling}

In \S 3.2.2, we adopted a disk-averaged dust temperature that scaled weakly 
with the host star luminosity, $\langle T_d \rangle \approx 25 \,
(L_{\ast}/L_{\odot})^{1/4}$\,K, as a basic assumption in the conversion of 
mm-wave continuum luminosities to dust disk masses (see Eq.~2).  From a 
theoretical perspective, this approximate scaling is appropriate only for the 
strictly optically thin case.  However, the disk midplanes responsible for the 
mm-wave emission studied here are heated indirectly, by radiation from an 
optically thick layer of dust in the disk atmosphere 
\citep[e.g.,][]{chiang97,dalessio98}.  So, there is a valid question about 
whether or not this intermediate step in the supply chain of thermal energy to 
the disk interior preserves the scaling with the original irradiation source 
properties (i.e., $L_{\ast}$).  There is no simple analytic prescription for 
answering this question a priori, and (to our knowledge) this has not been 
specifically addressed with a simple parameter study that uses continuum 
radiative transfer calculations in the literature.  So, to assess the basic 
applicability of this assumed scaling, we performed such a parameter study 
here.

This demonstration followed the general modeling formalism outlined by 
\citet{andrews09,andrews11}.  A two-dimensional model grid of dust densities 
was constructed from simple radial prescriptions for the surface densities, 
$\Sigma \propto r^{-1}$, and vertical scale heights, $H \propto r^{1.15}$, 
between an inner radius \citep[set by the assumed dust destruction temperature 
of 1500\,K; see][]{dullemond01} and a fixed outer edge at $r = 200$\,AU 
\citep[e.g.,][]{aw07a}, such that $\rho(r,z) = (\Sigma/\sqrt{2\pi}H) 
\exp{[-(z/H)^2/2]}$.  We adopted the settled dust prescription and opacities 
discussed by \citet{andrews11}, where small grains representing only 10\%\ of 
the total dust mass are distributed at larger vertical heights to intercept 
stellar irradiation, and the remaining 90\%\ of the mass is concentrated in 
larger grains near the disk midplane (with a scale height of only 0.2$H$).  We 
assumed $H \approx 28$\,AU at the outer edge in all models.  Three different 
scenarios for normalizing $\Sigma$ were considered: (1) an optically thin 
reference case where $M_d = 0.001$\,$M_{\odot}$; (2) a more massive counterpart 
where $M_d = 0.01$\,$M_{\odot}$; and (3) an a posteriori check on the 
regression results of \S 3.2.2, where $M_d/M_{\ast} \approx 0.4$\%.  We used 
the Monte Carlo radiative transfer code {\tt RADMC-3D} (v0.35; 
C.~P.~Dullemond)\footnote{\url{http://www.ita.uni-heidelberg.de/$\sim$dullemond/software/radmc-3d/}} to compute a two-dimensional dust temperature structure 
that is physically consistent with the irradiation of a given parametric 
density structure by an appropriate stellar photosphere model.  For each 
scenario, this was repeated for 25 different input stellar photosphere spectra 
interpolated from the {\tt NextGen} catalog as described in Appendix B, 
corresponding to the basic stellar properties along the 2.5\,Myr isochrone in 
the SDF00 models. 

\begin{figure}[t!]
\epsscale{0.5}
\figurenum{17}
\plotone{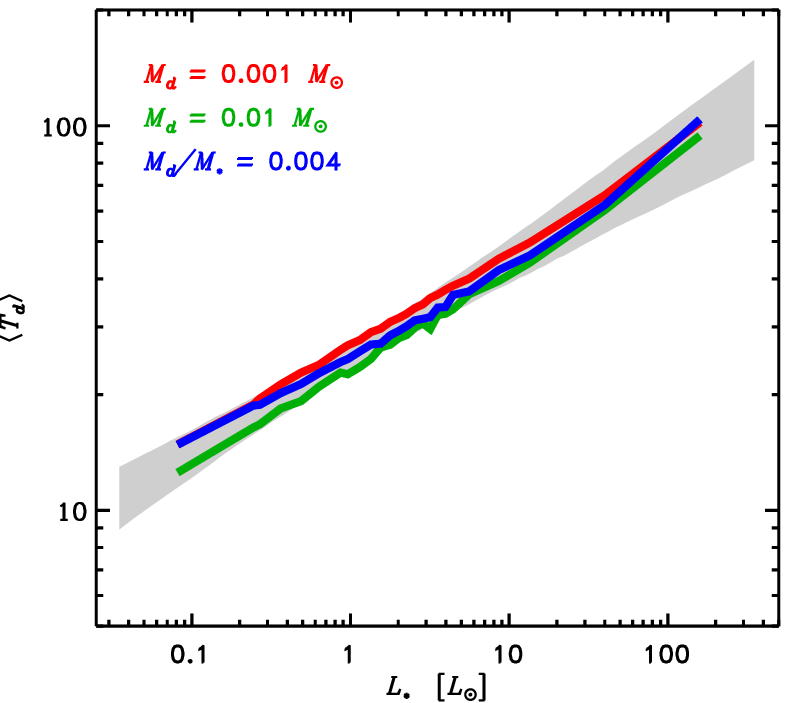}
\figcaption{A demonstration that the assumed scaling between the disk-averaged 
dust temperature (weighted by mass) and the stellar host luminosity is 
reasonable, using radiative transfer calculations for representative disk 
structure models.  The red and green curves show how $\langle T_d \rangle$ 
varies with $L_{\ast}$ for model disk structures with fixed masses of $M_d = 
0.001$ and 0.01\,$M_{\odot}$, respectively, and stellar parameters along the 
SDF00 2.5\,Myr isochrone.  The blue curve marks the analogous behavior for a 
linear $M_d \propto M_{\ast}$ scaling with the normalization found in \S 3.2.2, 
as an a posteriori verification of the assumed dust temperatures.  The gray 
shaded band encloses the assumed scaling of $\langle T_d \rangle = 
25\,(L_{\ast}/L_{\odot})^{1/4}$\,K, within a range of $\pm$2\,K in the 
normalization and $\pm$0.05 in the index, for reference.\label{fig:RT}}
\end{figure}

The results of these calculations are presented in Figure \ref{fig:RT}, where 
we show the mass-weighted average midplane temperature as a function of the 
input $L_{\ast}$ for Scenarios 1, 2, and 3 in red, green, and blue, 
respectively (the same results are found if we instead use the $\langle T_d 
\rangle$ that would be inferred from an inversion of Eq.~2, based on the input 
$M_d$ values and synthetic 1.3\,mm flux densities generated by {\tt 
RADMC-3D}).  The shaded gray region corresponds to the adopted dust temperature 
scaling, with an allowed range of $\pm$2\,K in the normalization and $\pm$0.05 
in the power-law index (for illustrative purposes).  Overall, this simple, 
controlled parameter study provides some basic verification that the $\langle 
T_d \rangle$ scaling with $L_{\ast}$ that was assumed in \S 3.2.2 is 
quantitatively reasonable for normal, representative disk parameters.  However, 
some caution should be exercised in generalizing or extrapolating the 
application of this behavior: variations in the fixed parameters -- 
particularly the scale height distribution -- can also induce changes in the 
shape of this scaling if they depend on the stellar parameters, as was 
described in more detail in \S 3.2.3.

\clearpage

\begin{deluxetable}{llr}
\tablecolumns{3}
\tablewidth{0pt}
\tabletypesize{\scriptsize}
\tablecaption{New 1.3\,mm Photometry Measurements\label{tab:obs}}
\tablehead{
\colhead{UT Date} & \colhead{Name} & \colhead{$F_{\nu}$ (mJy)\tablenotemark{a}} }
\startdata
2012 Nov 17 & IRAS 04187+1927   & $< 6.3$       \\
 & MHO 6             & $13.5\pm2.0$  \\
 & J04322415+2251083 & $< 6.0$       \\
 & J04333278+1800436 & $11.0\pm1.9$  \\
 & IRAS 04303+2240   & $< 6.0$       \\
 & J04333905+2227207 & $31.0\pm2.0$  \\
 & J04334171+1750402 & $6.0\pm2.0$   \\
 & J04390163+2336029 & $< 6.6$       \\
\enddata
\tablenotetext{a}{Upper limits are listed at the 3-$\sigma$ level.}
\tablenotetext{b}{Additional observations of these same targets were made on 
2012 Dec 2 and 2012 Dec 6, because the atmospheric opacity at 225\,GHz was too 
high to meet the desired RMS noise level in an individual track.  The data from 
all tracks were combined to measure these flux densities.  The three detections 
were significant and consistent in each of the three tracks.}
\tablenotetext{c}{As above, but additional observations were performed on 2013 
Feb 8 and combined.}
\end{deluxetable}

\clearpage

\begin{deluxetable}{llcccc}
\tablecolumns{6}
\tablewidth{0pt}
\tabletypesize{\scriptsize}
\tablecaption{Millimeter-wave Continuum Photometry Catalog\label{tab:Fmm}}
\tablehead{
\colhead{Name} & \colhead{Spectral Type\tablenotemark{a}} & \colhead{Ref.\tablenotemark{b}} & \colhead{$F_{\rm 0.89mm}$ (Jy)\tablenotemark{c}} & \colhead{$F_{\rm 1.3mm}$ (Jy)\tablenotemark{c}} & \colhead{Notes\tablenotemark{d}}}
\startdata
IRAS 04108+2910       & M0                 & 1       & $<0.0517$         & $<0.0198$         & e, m   \\
J04141188+2811535     & M6.25              & 1       & $<0.0509$         & $<0.0195$         & e, m   \\
FM Tau                & M0                 & 1       & $0.0288\pm0.0091$ & $0.0131\pm0.0027$ & m, e   \\
FN Tau                & M5                 & 1       & $0.0365\pm0.0050$ & $0.0168\pm0.0021$ & m, e   \\
CW Tau                & K3                 & 1       & $0.1601\pm0.0268$ & $0.0523\pm0.0070$ & m, m   \\
CIDA 1                & M5.5               & 1       & $0.0353\pm0.0174$ & $0.0135\pm0.0028$ & e, m   \\
MHO 1                 & M2.5               & 1       & $0.5295\pm0.1207$ & $0.2493\pm0.0311$ & e, m   \\
MHO 2 AB              & M2.5, M5.5$\pm$3   & 1, est  & $0.3017\pm0.0555$ & $0.1377\pm0.0142$ & e, m   \\
\enddata
\tablenotetext{a}{Unless otherwise specified, the assumed spectral type 
uncertainty is $\pm$1 subclass.}
\tablenotetext{b}{The spectral classifications were collated from the following 
references: 1 = \citet{luhman10}, 2 = \citet{hartigan03}, 3 = 
\citet{hernandez04}, 4 = \citet{furlan11}, 5 = \citet{white01}, 6 = 
\citet{itoh05}, 7 = \citet{espaillat10}, 8 = \citet{duchene99}, 9 = 
\citet{kraus09}, 10 = \citet{white99}, 11 = \citet{prato02}, 12 = 
\citet{schaefer12}, 13 = \citet{hartigan94}, 14 = \citet{mathieu97}, 15 = 
\citet{white05}, 16 = \citet{dahm11}, 17 = \citet{roberge01}, 18 = 
\citet{massarotti05}, 19 = \citet{mora01}, 20 = \citet{gahm08}.  The `est' flag 
means that the classification was crudely estimated based on the type of its 
primary companion and optical/near-infrared contrast ratios: more details on 
this modification to the classification process are available in Appendix B.  
The classifications for FT Tau, ITG 1, and IRAS 04370+2559 are unknown: we 
adopt the wide spectral type bins advocated by \citet{luhman10}, listed here in 
parenthesis.}
\tablenotetext{c}{Upper limits are listed at the 3-$\sigma$ level.}
\tablenotetext{d}{The `m' and `e' flags denote whether each source has a flux 
density measurement available at (or near) 0.89 and 1.3\,mm: `m' implies direct 
measurement and `e' implies an extrapolation as described in the text.  Notes 
in parenthesis indicate unresolved multiple systems where composite 
measurements were applied to the primaries, and upper limits to the secondaries 
(see \S 3.1).}
\tablenotetext{e}{The companions V807 Tau B (itself a binary) and V710 Tau B 
have no observational disk signatures, and are therefore excluded from this 
Class II sample.  The V807 Tau system has no resolved mm-wave measurement, but 
we associate all of the composite emission with the primary component.  
\citet{harris12} found no mm-wave emission associated with V710 Tau B.}
\end{deluxetable}

\clearpage

\begin{deluxetable}{l|cccc|cccc|cccc}
\rotate
\tablecolumns{13}
\tablewidth{0pt}
\tabletypesize{\tiny}
\tablecaption{Stellar Host Masses and Ages \label{tab:stars}}
\tablehead{
\colhead{} & \multicolumn{4}{c}{DM97} & \multicolumn{4}{c}{BCAH98} & \multicolumn{4}{c}{SDF00} \\
\colhead{Name} & \colhead{$\log{\,M_{\ast}}$} & \colhead{CI$(\log{\,M_{\ast}})$} & \colhead{$\log{\,t_{\ast}}$} & \colhead{CI$(\log{\,t_{\ast}})$} & \colhead{$\log{\,M_{\ast}}$} & \colhead{CI$(\log{\,M_{\ast}})$} & \colhead{$\log{\,t_{\ast}}$} & \colhead{CI$(\log{\,t_{\ast}})$} & \colhead{$\log{\,M_{\ast}}$} & \colhead{CI$(\log{\,M_{\ast}})$} & \colhead{$\log{\,t_{\ast}}$} & \colhead{CI$(\log{\,t_{\ast}})$}
}
\startdata
IRAS 04108+2910       & -0.22 & (-0.44, -0.18) & 0.80  & (0.13, 1.23)   & -0.14 & (-0.27, -0.05) & 0.99  & (0.48, 1.40)   & -0.23 & (-0.37, -0.15) & 0.95  & (0.49, 1.39)   \\
J04141188+2811535     & -0.93 & (-1.23, -0.83) & 0.82  & (0.50, 1.17)   & -1.32 & (-1.41, -1.02) & 0.60  & (-1.12, 0.81)  & -1.08 & (-1.34, -0.95) & 0.81  & (0.17, 1.12)   \\
FM Tau                & -0.33 & (-0.50, -0.20) & 0.15  & (-0.18, 0.62)  & -0.05 & (-0.22, -0.01) & 0.67  & (0.29, 1.00)   & -0.22 & (-0.34, -0.11) & 0.53  & (0.27, 0.92)   \\ 
FN Tau                & -0.92 & (-0.96, -0.79) & -1.57 & (-2.41, -1.23) & -0.64 & (-0.79, -0.45) & -1.30 & (-1.93, -0.81) & -0.63 & (-0.76, -0.57) & -1.11 & (-1.55, -0.47) \\ 
CW Tau                & 0.02  & (-0.13, 0.10)  & -0.13 & (-0.50, 0.27)  & 0.30  & (0.21, 0.36)   & 0.46  & (0.11, 0.80)   & 0.20  & (0.13, 0.27)   & 0.35  & (0.06, 0.68)   \\
CIDA 1                & -0.87 & (-0.93, -0.76) & 0.15  & (-0.52, 0.25)  & -1.04 & (-1.08, -0.74) & 0.10  & (-2.09, 0.17)  & -0.89 & (-0.98, -0.75) & 0.43  & (-0.36, 0.48)  \\
MHO 1                 & -0.71 & (-0.76, -0.48) & -0.10 & (-2.27, 0.54)  & -0.20 & (-0.40, -0.10) & -0.86 & (-2.15, 0.46)  & -0.45 & (-0.54, -0.36) & -2.01 & (-2.40, 0.41)  \\
MHO 2 AB\tablenotemark{a} & -0.49 & (-0.54, -0.25) & -1.17 & (-1.78, -0.50) & -0.09 & (-0.21, 0.05)  & -0.56 & (-1.22, -0.06) & -0.32 & (-0.38, -0.16) & -0.09 & (-1.28, -0.08) \\
\enddata
\tablenotetext{a}{The $\log{M_{\ast}}$ values (and their confidence intervals) 
in these composite multiple systems were determined by properly combining the 
marginal probability density functions determined for their individual 
components, as described in the text.  The corresponding $\log{t_{\ast}}$ 
values (and their confidence intervals) correspond to the primary components.}
\tablenotetext{b}{These systems are either known or strongly suspected to be 
associated with nearly edge-on disks.  Since such a viewing geometry 
substantially obscures the starlight, the derived estimates of $L_{\ast}$ for 
these cases are not used to determine masses and ages.  Instead, we compute the 
weighted mean and standard deviation of the luminosities for the ensemble of 
sources with the same spectral type ($\pm$1 subclass), and associate them with 
$L_{\ast}$ and $\sigma_L$ in the Bayesian inference method for \{$M_{\ast}$, 
$t_{\ast}$\}: details are available in Appendix B.}
\tablecomments{The units of the compiled stellar masses and ages are in units 
of $M_{\odot}$ and Myr, respectively.  The quantity CI($X$) is meant to 
represent the extreme values of the variable $X$ that span its 68.2\%\ 
(``1-$\sigma$") confidence interval, derived from its associated marginal 
probability density function (see \S 3.2.1).  As an example, consider the 
results for CIDA 12 from the SDF00 grid: an alternative (perhaps more familiar) 
way of quoting the measurements would be $\log{M_{\ast}} = 
-0.65^{+0.10}_{-0.18}$ and $\log{t_{\ast}} = 0.67^{+0.28}_{-0.15}$. } 
\end{deluxetable}

\clearpage

\begin{deluxetable}{lcrr}
\tablecolumns{4}
\tablewidth{0pt}
\tabletypesize{\scriptsize}
\tablecaption{Stellar Spectrum Fitting Results\label{tab:HR}}
\tablehead{
\colhead{Name} & \colhead{$\log{(T_{\ast}/{\rm K})}$} & \colhead{$\log{(L_{\ast}/L_{\odot})}$} & \colhead{$A_V$ (mags)}}
\startdata
IRAS 04108+2910       & $3.5855\pm0.0237$ & $-0.691\pm0.290$ & $1.73\pm1.09$    \\
J04141188+2811535     & $3.4717\pm0.0189$ & $-1.746\pm0.082$ & $0.00\pm1.43$    \\
FM Tau                & $3.5855\pm0.0237$ & $-0.393\pm0.122$ & $1.94\pm0.64$    \\
FN Tau                & $3.4948\pm0.0202$ & $-0.082\pm0.097$ & $2.02\pm0.73$    \\
CW Tau                & $3.6749\pm0.0156$ & $0.384\pm0.205$  & $3.17\pm1.20$    \\
CIDA 1                & $3.4854\pm0.0199$ & $-0.901\pm0.089$ & $2.06\pm0.68$    \\
MHO 1                 & $3.5425\pm0.0181$ & $0.230\pm0.883$  & $12.75\pm3.28$   \\
MHO 2 A\tablenotemark{a} & $3.5425\pm0.0181$ & $0.019\pm0.189$  & $9.00\pm1.08$  \\
\enddata
\tablenotetext{a}{The stellar parameters for these close pairs were determined 
by simultaneous fits to both their composite SEDs and any available contrast 
ratios, assuming that both components have the same extinction.}
\tablenotetext{b}{These stars are known or suspected to host edge-on disks.  
The luminosities and extinctions reported here are the results of the formal 
measurements described in Appendix B.  However, their $L_{\ast}$ values are 
modified before use in \S 3.2, as described in the text.}
\tablenotetext{c}{The components V807 Tau Bab and V710 Tau B are included here 
for completeness, but are not considered part of the Class II sample.  The 
stellar photosphere parameters listed here were used to subtract off their 
contribution to the composite SEDs shown in Appendix B.}
\end{deluxetable}

\clearpage

\bibliography{references}

\end{document}